\documentstyle[epsf,epsfig]{report}
\newcommand \beq{\begin{eqnarray}}
\newcommand \eeq{\end{eqnarray}}
\newcommand \be{\begin{eqnarray}}
\newcommand \ee{\end{eqnarray}}
\newcommand{\set}[2]{\newcommand{#1}{#2}}
\set{\pa}{\partial \over \partial\, }
\set{\leftvector}{\stackrel{\leftarrow}{\partial }}
\set{\rightvector}{\stackrel{\rightarrow}{\partial }}
\set{\ba}{\bar }
\set{\e}{\epsilon }
\set{\intp}{\int{dp\over (2 \pi)^3}}
\set{\ppl}{(p+\frac q 2)}
\set{\pmi}{(p-\frac q 2)}
\begin{document}
\title{Collective modes in asymmetric nuclei}
\author{Klaus Morawetz$^a$, Uwe Fuhrmann$^{b}$, Rainer
  Walke$^{b,c}$
\\
$^a$ LPC-ISMRA, Bld Marechal Juin, 14050 Caen
 and  GANIL,
Bld Becquerel, 14076 Caen, France\\
$^b$ Fachbereich Physik, Universit\"at Rostock, 18051 Rostock,
Germany\\
$^c$current address: Max-Planck-Institute for
    demographic research, Rostock, Germany
}
\maketitle
\begin{abstract}
The collective motion of a finite nuclear system is investigated by
numerical simulation and by linear response theory. Using a pseudo-particle simulation
technique we analyze the giant resonances with a multipole decomposition scheme.
We examine the energy and the damping of different
giant collective modes and
obtain the dependence of these quantities on the proton--neutron ratio.
In simulations of finite nuclei including only mean fields the
centroid energy of the resonance decreases with higher asymmetry due
to the
change of the compressibility and the damping increases. The
collisional correlation in turn leads to a decreasing damping with
increasing asymmetry.

Alternatively the giant collective modes in asymmetric nuclear matter
are investigated within linear response theory including the
collisional correlations via a dynamic 
relaxation time approximation.
For a multicomponent system we derive a coupled dispersion relation and show that two sources
of coupling appear: (i) a coupling of isoscalar and isovector
modes due to the action of different mean-fields and (ii) an explicit new
coupling in asymmetric matter due to collisional interaction. We
show that the latter is responsible for a new mode arising
besides isovector and isoscalar modes. The comparison with simulation
results as well as experiments is performed. It is stressed that the
surface effects and the collisional correlations are both essential in
order to
describe the correct damping behavior. A model is presented which
allows to combine surface and collisional effects. Higher-order modes
like the isoscalar dipole mode which where recently measured are
discussed within this frame.

Collective motion beyond the linear regime is demonstrated for the
example of 
large-amplitude isoscalar giant octupole excitations
in finite nuclear systems. Depending on the 
initial conditions we observe either clear octupole modes or
over-damped octupole modes which
decay immediately into quadrupole ones. This dependence on initial
correlations represents a behaviour beyond 
linear response.

\end{abstract}

\tableofcontents
\chapter{Introduction}
Giant resonances are currently enjoying a resurgence of attention as a
tool for the investigation of many 
particle effects in
finite quantum systems.
The study of giant resonances offers the possibility to learn more
about nuclear forces. Of special interest are modern experiments with
exotic nuclei which
broaden our knowledge in a new degree of 
freedom: the variability in the proton to neutron ratio.
The treatment of collective modes in nuclear matter is documented in an 
enormous literature starting from their discovery \cite{BoGe37} up to
recent discussions
\cite{BBB83}
-\cite{LDH96} 
and citations therein.
 
We want to investigate the collective excitations in asymmetric nuclear
matter \cite{FMW98,CLN97} by numerical simulation of the Vlasov equation and
by comparison with the linear response theory.
While the former leads to an insight into finite-size effects like
surface, the latter allows to consider collisional correlations in
a control-able way. Both methods are contrasted and we demonstrate
that the collisional correlations as well as the surface effects are
important to describe the experimental damping of giant resonances.

The damping mechanisms of collective motions in excited nuclei
are a topic of continuing debate. 
Mainly two lines of thought are pursued.
In one line of thought it is assumed that collisions are the only physical reason for 
damping which is described via a Fermi liquid approach 
\cite{FMW98},\cite{KAM69}-
\cite{GTO98,TKL99}.
The other line of thought considers new features of the finite nucleus,    
such as
surface oscillations and a level
density with finite spacing.  The investigations are performed without 
inertia \cite{SHB75}-
\cite{KIN98,LCA98}
or by including inertia \cite{OBB97},\cite{ALB89}-
\cite{LDH96}; note that 
inertia is absent in infinite matter. 

Both classes of models predict a comparable degree of damping necessary
to reproduce the experimental data. Consequently, it is an open question which is the 
correct physical reason for damping.
Of course, the correct
description has to assume a finite nucleus consisting of nucleons which are
bound via the mean field, through which the nucleons undergo mutual collisions and 
where the surface is formed by the particles themselves.
These features are principally included in Boltzmann-Uehling-Uhlenbeck-(BUU) 
simulations 
\cite{CTG93,BAR96,MWS98} or in its nonlocal extensions \cite{SLM96,MLSCN98}. In full 
simulations, however,  we will not gain a simple insight into the physical origins of 
the damping mechanism, in particular, how much is contributed by the surface and how 
much by collisions.

One aim of this article is therefore to compare both pictures
in the frame of linear response theory. 
Within the collision-free Vlasov equation the linear response of
finite systems is well known \cite{BDT86} and allows one to calculate
the strength function of finite nuclei. The resulting damping, however, does not
reproduce the experimental damping of giant resonances since
collisions are absent. This motivates us to develop a linear response
theory including collisions.

While most of the theoretical treatments of oscillations rely on the linear response 
method or RPA methods, large amplitude oscillations require methods
beyond this level.  
In particular the question of the appearance of chaos has
recently been investigated \cite{BGZS94,schuckbaldo,Mr97}. 
The hypothesis was established that the octupole mode is over-damped due to negative 
curved surface and consequent additional chaotic damping
\cite{JarSwi93,BloShi93,BloSka97}. Here we want to discuss in which
conditions one might observe octupole modes at least in Vlasov -
simulations of giant resonances which will turn out to be dependent on
initial conditions and are consequently an effect beyond linear response.

The outline is as follows: In the second chapter we will give the numerical results of
pseudo-particle simulation of the Vlasov equation. This will provide us
with some insight into the magnitude of finite size effects on the
damping. In the third chapter we present the linear response
result of the Vlasov equation including collisional
correlations. We will consider a simplified picture of infinite matter
response and will consider the Steinwedel Jensen
picture. We discuss two possibilities to include
surface contributions in the linear response formalism. This allows us
to describe the experimental temperature dependence of the damping of giant dipole
resonances as well as the structure function of the isovector dipole
resonances. The surface consideration allows to describe the isoscalar
dipole resonance as a higher order mode. The comparison between linear response results and
simulations is performed and finally we discuss nonlinear
effects beyond linear response for the example of giant octupole modes.

\chapter{Pseudoparticle simulation}
We will describe the giant resonance first by a kinetic equation using
a pseudo-particle simulation \cite{BD88}. 
The kinetic equation for the quasiclassical distribution function
reads for neutrons (for protons analogously)
\be
&&{\dot f}_n({\bf p,r},t)+\frac{{\bf p}}{m}{\bf \partial_r}
f_n({\bf p,r},t)-{\bf \partial_r}(U_n+U_{\rm ext}) {\bf \partial_p}f_n({\bf p,r},t)=I_{\rm corr}
                      \label{klassVlasov}
\nonumber\\&&
\ee
with the collisional integral $I_{\rm corr}$ discussed later and the self-consistent mean-field potential $U$ given by a schematic Skyrme 
type \cite{ColDiT95}  
  \begin{eqnarray} 
    U_{n/p}(\varrho,I) &=& a ({\textstyle \frac{\varrho}{\varrho_0}}) 
      + b ({\textstyle \frac{\varrho}{\varrho_0}})^s
      \pm c I ({\textstyle \frac{\varrho}{\varrho_0}})
\label{hf}
  \end{eqnarray}
with $a=-356{\rm MeV}$, $b=303{\rm MeV}$ , $c=54{\rm MeV}$ and $s=7/6$.
Here the neutron excess is 
$I=\frac{\varrho_n-\varrho_p}{\varrho_n+\varrho_p}$, the neutrons feel the
+--sign potential, the protons the opposite one.

By multiplying the kinetic equation by $1,p$ or $E={p^2\over 2m}+U$
respectively one obtains the
balance for particle density $\rho$, momentum density $u$ and energy density ${\cal E}$. 
Since the collision
integrals vanish for density and momentum balance we get the usual
balance equations
\be
&&{\partial \rho({\bf r},t)\over \partial t}+{\bf {\partial \over \partial r}} \int
{d {\bf p}\over (2 \pi \hbar)^3} {\partial E\over \partial {\bf p}}  f({\bf p,r},t)=0\nonumber\\
&&{\partial u_i({\bf r},t)\over \partial t} +{\partial \over \partial r_j} \int
{d{\bf p}\over (2 \pi \hbar)^3} (p_i {\partial E\over \partial p_j} f({\bf p,r},t)+{\cal
    E}({\bf r},t)\delta_{ij})=0
\ee
where the mean field energy of the system varies as 
\be
\delta {\cal E}&=&\int
{d {\bf p}\over (2 \pi \hbar)^3} {\delta {\cal E}\over \delta
    f({\bf p,r},t)} \delta f({\bf p,r},t)
\nonumber\\
&=&\int
{dp\over (2 \pi \hbar)^3} ({p^2\over 2 m}+U) \delta f({\bf p,r},t)
\ee
such that from (\ref{hf}) follows the total energy density as
\be
{\cal E}({\bf r},t)=\int{d {\bf p}\over (2\pi \hbar)^3} {p^2\over 2 m} f({\bf p,r},t)+{a\over 2} ({\textstyle \frac{\varrho^2({\bf r},t)}{\varrho_0}}) 
      + {b \rho({\bf r},t) \over s+1} ({\textstyle \frac{\varrho({\bf r},t)}{\varrho_0}})^{s+1}
      \pm {c\over 2} I ({\textstyle \frac{\varrho^2({\bf r},t)}{\varrho_0}}).
\label{en}
\ee
With the help of this quantity the balance of energy density reads
from (\ref{klassVlasov})
\be
&&{\partial {\cal E}({\bf r},t) \over \partial t} +{\bf {\partial \over \partial r}} \int
{d{\bf p}\over (2 \pi \hbar)^3} E {\partial E\over \partial {\bf p}}
f({\bf p,r},t)=-{\partial \over \partial t}E_{\rm
    corr}({\bf r},t)
\label{e1}
\ee
with the correlation energy \cite{M94,MK97} arising from the
collisional side. The
collisional side we will consider later in linear response. Here for
the simulation we will neglect the collisions and will restrict
ourselves to the
Vlasov kinetic equation. In this way we will learn what are the
effects of finite size and what are the effects of collisions.

\begin{figure}[h]
\parbox[t]{19cm}{
\parbox[l]{7.3cm}{
\psfig{file=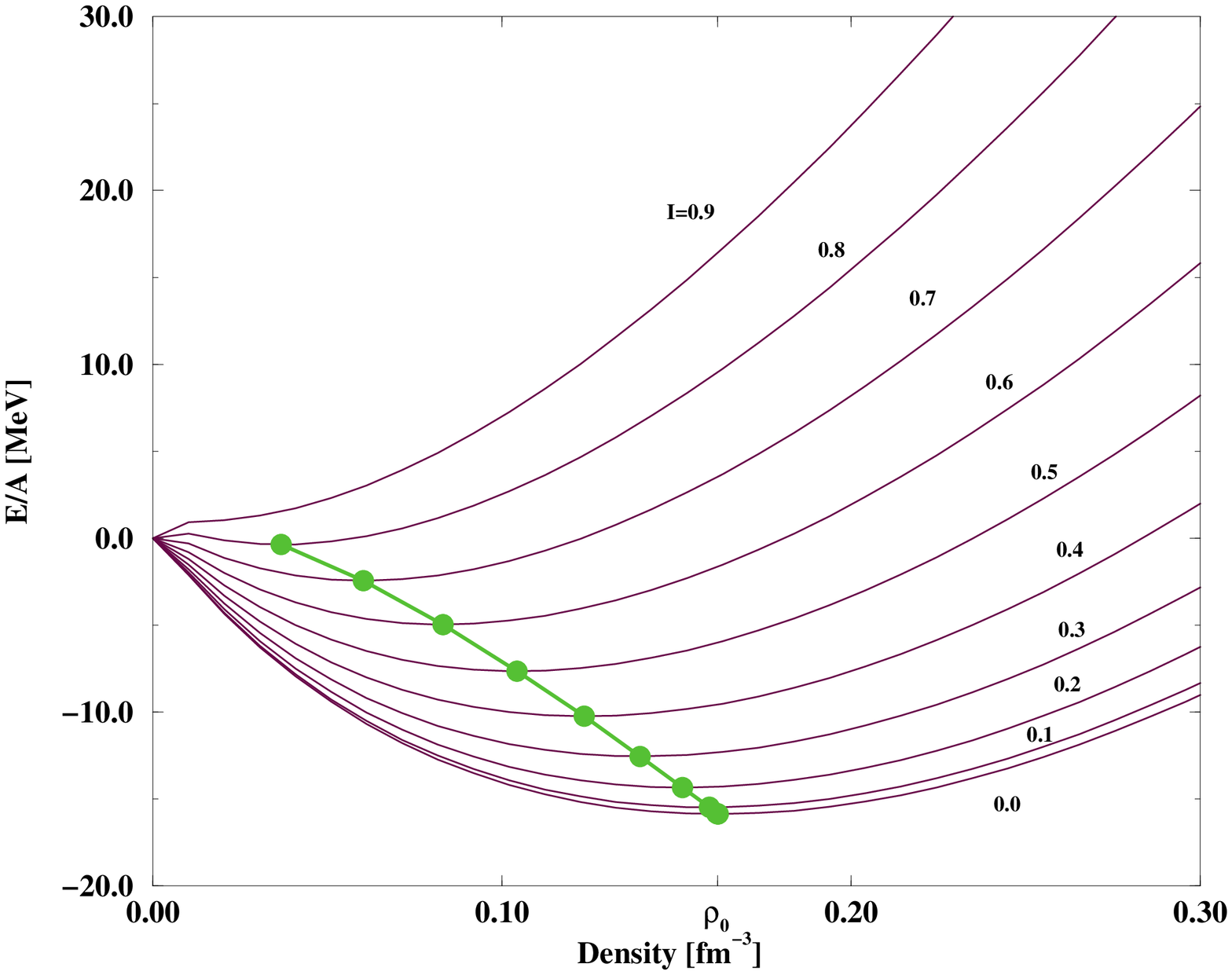,width=6cm,height=7cm,angle=-90}
}
\parbox[c]{5.4cm}{
\psfig{file=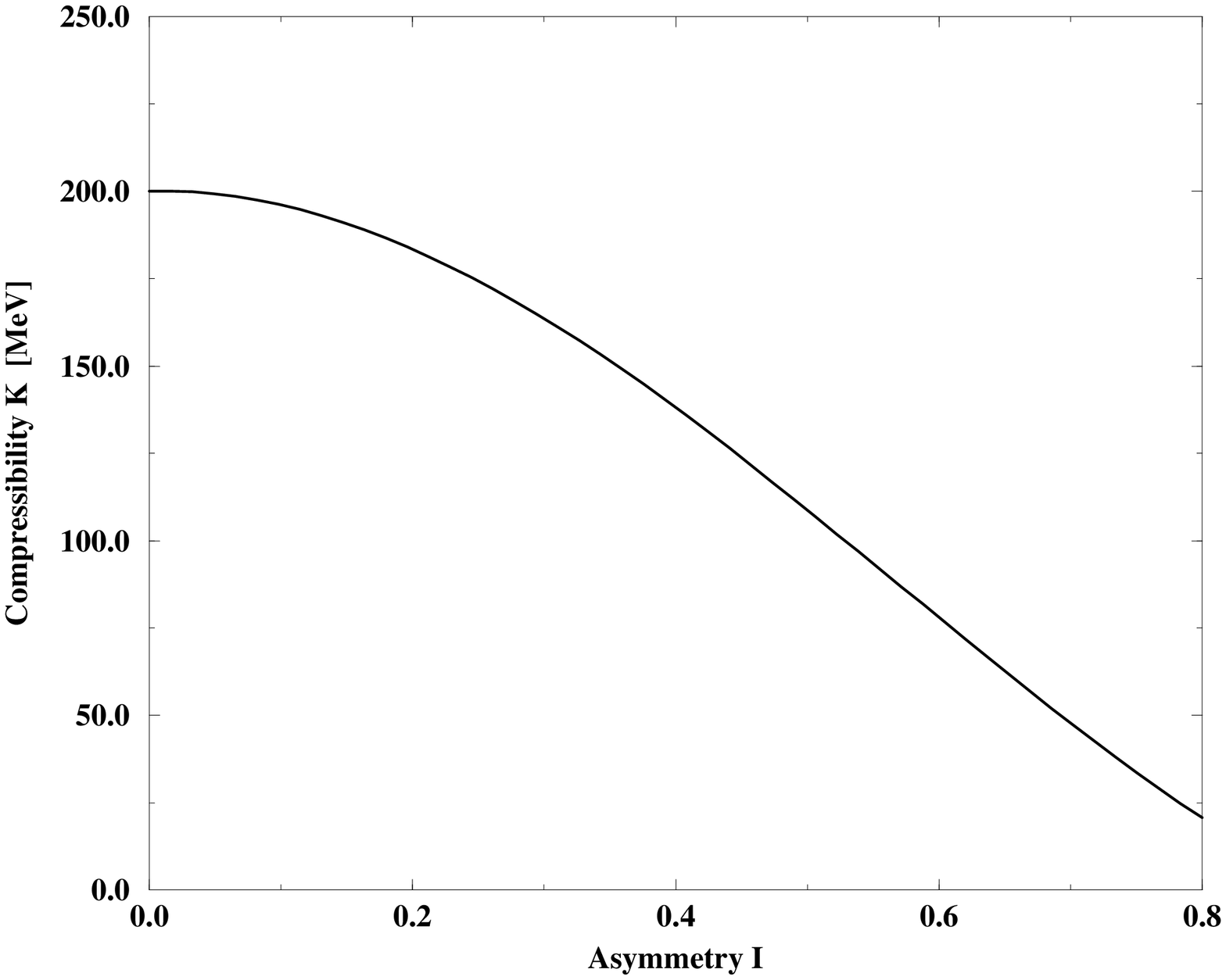,width=6cm,height=5.4cm,angle=-90}
}
\hspace{0.3cm}\parbox[r]{5cm}{
\caption{Left: Mass number dependence of the energy for different neutron excess
$I=\frac{\varrho_n-\varrho_p}{\varrho_n+\varrho_p}$.
The dots denote the saturation densities.
With increasing asymmetry the saturation density and the binding
energy decrease. Right: Asymmetry dependence of compressibility modulus.
With increasing asymmetry this modulus decreases. 
\label{pot54}}
}}
\end{figure}

Let us now first look at the input for the simulation. The mass number dependence of binding energy ${\cal E}/A$ from
(\ref{en}) is shown in 
figure \ref{pot54}.  One sees that with increasing asymmetry the binding energy 
becomes weaker and consequently the compressibility decreases.
The compression modulus is defined as the derivative of the Energy per
particle
\be
K=9 n^2 {\partial ({\cal E}/A)\over \partial n^2} 
\ee
and took for $I=0$ the value $K=200\; {\rm MeV}$ according to (\ref{en}). Figure \ref{pot54} shows the asymmetry dependence 
of K. One sees that for neutron matter no binding energy would occur and a very low 
compressibility. 

We solve now the kinetic equation by representing the distribution function $f({\bf 
p,r},t)$ by a sum of
N pseudo-particle distributions
\begin{eqnarray}
f({\bf p,r},t) \approx f_0({\bf p,r},t) &=& \sum\limits_{i=1}^{A N} 
      \frac{1}{N} f_S({\bf p}-{\bf p_i}(t),{\bf r}-{\bf r_i}(t)) \; 
\label{222}
\end{eqnarray}
and use Gaussian pseudo-particles 
\begin{eqnarray}
f_{S}({\bf p-p_1,r-r_1}) = c\,\; e^{-({\bf p-p_1})^2/2\sigma_p^2}\;\; e^{-({\bf 
r-r_1})^2/2\sigma_r^2}
\label{33}
\end{eqnarray} 
at ${\bf r_1}$ with momentum ${\bf p_1}$ \cite{W82}.
These pseudo-particles follow 
classical Hamilton equations
\begin{eqnarray}
\dot{\bf p}_i&=&-{\bf \nabla} U,\; \; \; \dot{\bf r}_i=\frac{{\bf p}_i}{m} 
\; .
\end{eqnarray}
The only fluctuations introduced are due to some small unavoidable numerical noise\cite{ColBur93}.
We are using  300
pseudo-particles per nucleon and a pseudo-particle width of
 $\sigma_r = 0.466 \;{\rm fm}$. They are adjusted both in such a way that the
 experimental energy of the $^{40}Ca$ giant monopole mode is reproduced. 
With these fixed two parameters the experimental behavior of centroid
energy with mass number is than reproduced over the full range for
giant monopole and giant dipole resonances. We have checked 
different numbers of test particles. The dependence of observables on the width is 
discussed in \cite{MP96}.
Numerically the ground states of nuclei are realized by Wood- Saxon shapes of density 
and Fermi spheres in momentum.

\section{Multipole analysis}

Provided we have now solved the kinetic equation we will have the
distribution function represented by $N$ pseudo-particles. Since we are
interested in moments of the distribution function, the density,
current and energy, we can expand these moments in the test particle
representation as well.
The momentum distribution of these moments can be obtained by spatial
integration: $F_a({\bf p})=\int d{\bf r} \,a\, f({\bf p,r})$ with, for the mass
distribution, $a=1$, for isospin, $a=\tau$, for kinetic energy,
$a=\frac{p^2}{2m}$, and for kinetic isospin energy,
$a=\frac{\tau p^2}{2m}$ respectively. Decomposed into spherical coordinates
they read 
\begin{eqnarray}
F_a(p,\vartheta,\varphi) &=& \sum_{i=1}^{A N} \frac{(2\pi)^3}{N}  a_i
  \frac{\delta(p-p_i)}{p_i^2} \delta(\varphi-\varphi_i)
  \frac{\delta(\vartheta-\vartheta_i)}{\sin(\vartheta_i)} \;.
\end{eqnarray}
Radial integration determines now a spherical distribution
\begin{eqnarray}\label{f1}
\bar{F}_a(\vartheta,\varphi) &=& \sum_{i=1}^{A N} \frac{1}{N}  a_i
  \delta(\varphi-\varphi_i) \frac{\delta(\vartheta-\vartheta_i)}
{\sin(\vartheta_i)}
\end{eqnarray} 
which can be decomposed into spherical harmonics
\begin{eqnarray}\label{f2}
\bar{F}_a(\vartheta,\varphi) &=& \sum_{l=0}^{\infty} \sum_{m=-l}^{l} 
  a_{lm} Y_{lm}(\vartheta,\varphi)\;, \\
 a_{lm} &=& \sum_{i=1}^{A N} \frac{1}{N}  a_i 
 Y_{lm}^\star(\vartheta_i,\varphi_i).
\end{eqnarray}
\begin{figure}[h]
\parbox[]{18cm}{
\parbox[]{10cm}{
\psfig{file=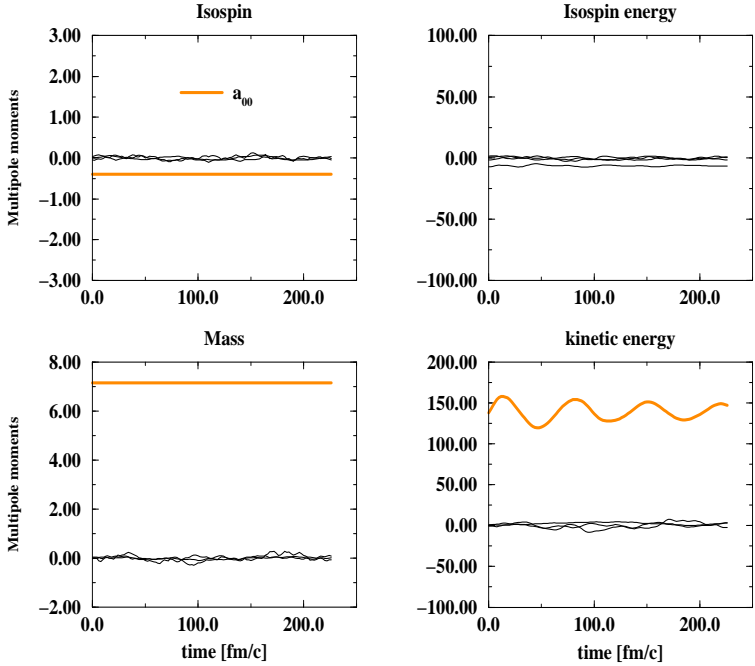,width=9cm,height=10cm,angle=-90}
}
\hspace{0.5cm}
\parbox[t]{7cm}{
\caption{The picture shows for isoscalar monopole oscillations of $^{90}_{40}Zr$ the time evolution of multipole 
  moments
  $\protect\sqrt{\frac{2l+1}{4\pi}} a_{l0}(t), l=0,1,2,3$
  corresponding to monopole, dipole, quadrupole and octupole
  oscillation for the
  distribution of isospin $a=\tau$, kinetic isospin energy
  $a=\frac{\tau p^2}{2m}$, mass $a=1$ and kinetic energy
$a=\frac{p^2}{2m}$. \label{b212}} 
}}
\end{figure}
The observable distributions $\bar{F}_a(\vartheta,\varphi)$ are normalized 
to $\sqrt{4\pi} \; a_{00}$, i.e. to mass number $A$, total isospin $T$,
kinetic energy $E_{kin}$ and kinetic isospin energy $E_{kinT}$, respectively. 
The polar angular distribution of moments reads now from
(\ref{f1}) and (\ref{f2})
\begin{eqnarray}
\hat{F}_a(\vartheta) = \sum_{i=1}^{A N} \frac{1}{N}  a_i
 \frac{\delta(\vartheta-\vartheta_i)}
{\sin(\vartheta_i)}\equiv \int_0^{2\pi} d\varphi \, 
\bar{F}_a(\vartheta,\varphi) &=&
\sum_{l=0}^{\infty} a_{l0} {\textstyle \sqrt{\frac{2l+1}{4\pi}}} 
P_l(\cos\vartheta)
\label{f3}
\end{eqnarray} 
with Legendre polynomials $P_l$.
As a measure for the strength of the resonances we obtain now the
coefficients $a_{l0}$ of the corresponding moment from (\ref{f3}) as
\be
a_{l0} \sqrt{\frac{2l+1}{4\pi}}={2 l +1 \over 2}\sum_{i=1}^{A N} \frac{1}{N}  a_i
\frac{P_l(\cos\vartheta_i)}
{\sin(\vartheta_i)}.
\ee
These amplitudes of multipole moments, $a_{l0} \sqrt{\frac{2l+1}{4\pi}}$, 
are displayed
as a function of time in figures \ref{b212},\ref{b213},\ref{b217} and \ref{b221}.
The value $a_{10}$ means the dipole moment, vanishing for isoscalar resonances,
$a_{20}$ is characterizing the quadrupole oscillations and $a_{30}$
the octupole ones, etc.

\section{Isoscalar monopoles}
The first analyzed mode is the isoscalar giant monopole mode which
plays an important role in the determination of nuclear
compressibility.
The connection of the compressibility and the energy
of giant monopole resonances is discussed e.g. in \cite{Blaizot80}.

The figure \ref{b212} shows simulation results for a monopole
oscillation. Here the excitation has been performed by adding an extra
momentum to the test-particles in the direction of the center of mass.
We have excited in this way a clear monopole breathing mode which can
be seen in the oscillation of the kinetic energy. The corresponding
mean field energy performs the opposite oscillations that the total
energy is constant. The fact that energy is oscillating between
kinetic and correlational ones describes why we have here a
compressional or breathing mode. All other modes remain unexcited. The
finite value of the isospin and mass for $l=0$ reflects the
conservation of isospin and particle number. 

\begin{figure}[h]
\parbox[]{18cm}{
\parbox[]{10cm}{
\psfig{file=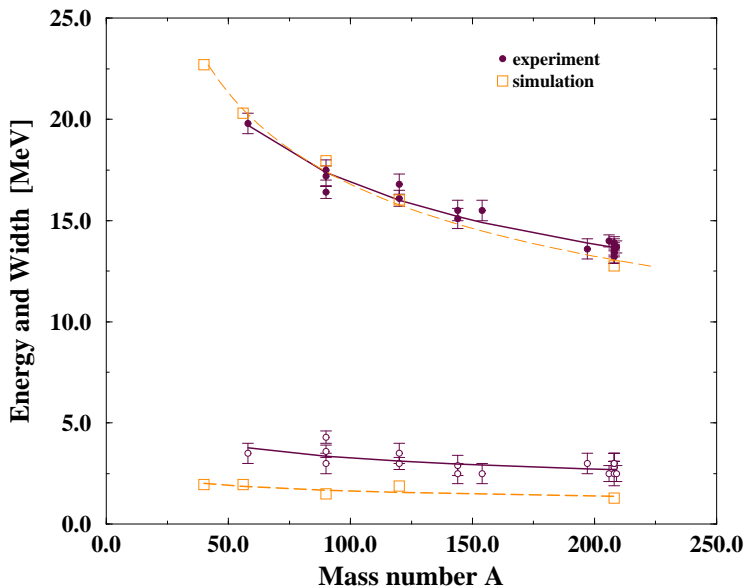,width=8cm,angle=-90}
}
\hspace{2.5cm}\parbox[t]{5cm}{
\caption{The mass number dependence of giant monopole 
oscillation energy and damping width. The solid line denotes the  experimental values 
while the 
dashed line describes the Vlasov simulation results.
\label{blaizot2}} 
}}
\end{figure}

With the previously chosen fixed  width of test-particles we see from figure 
\ref{blaizot2} that the experimental mass number dependence
of monopole oscillation is well reproduced over the whole range of
mass numbers. However, the experimental damping can be seen to be largely underestimated by the Vlasov 
simulation which yields about $2$ MeV. This is a first indication that
the mean field, even for a finite system, 
cannot account for the whole damping and dissipation. Instead we have to take into 
account collisional correlations which will be performed later.  

\begin{figure}[h]
\parbox[]{18cm}{
\parbox[]{10cm}{
\psfig{file=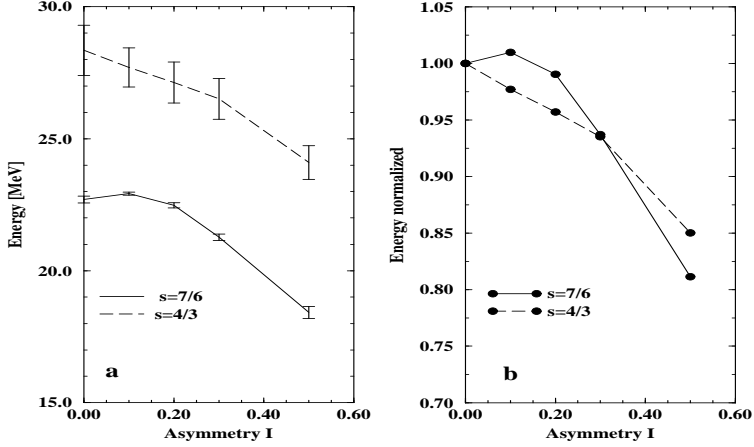,height=10cm,width=6cm,angle=-90}
}
\hspace{0.5cm}\parbox[]{7cm}{
\caption{The giant monopole energy (a) and normalized to the symmetric
  case (b) versus
asymmetry for two different parameterizations of the mean field
according to (\protect\ref{e1}). The stiff equation of 
state (black line) leads to higher monopole energy and shows a more linear decrease while the 
soft equation of state (grey line) shows a weak dependence for small
asymmetry and decreases faster for higher asymmetries.\label{ab7643}} 
}}
\end{figure}

It is now instructive to examine the isospin dependence of the isoscalar monopole. 
Since this is the experimental value which determines the
nuclear compressibility, the isospin dependence is of direct importance. 
In figure  \ref{ab7643} the dependence on asymmetry of the isoscalar monopole energy 
is plotted for a hard as well as a stiff equation of state. Of course,
the hard equation of state leads to a higher monopole energy
corresponding to a higher compressibility. 
Analogously to 
figure \ref{pot54} the monopole energy decreases with the asymmetry.
The hard equation of state leads to a more linear decrease
while the soft equation of state remains almost unchanged up to a
certain asymmetry and then decreases faster. 

\section{Isovector dipole resonances}

On the next figure \ref{b213} the multipole analysis for $^{90}_{40}Zr$ 
oscillations is performed. The excitation is now chosen as an isovector
dipole one created by a shift of proton against neutron spheres. One
recognizes that a clear dipole $a_{10}$ mode is excited which is seen
in the oscillation of the isospin and isospin energy. The kinetic
energy now remains constant in contrast to the isoscalar mode
because we have no compression mode. 
\begin{figure}[h]
\parbox[]{18cm}{
\parbox[]{11cm}{
\psfig{file=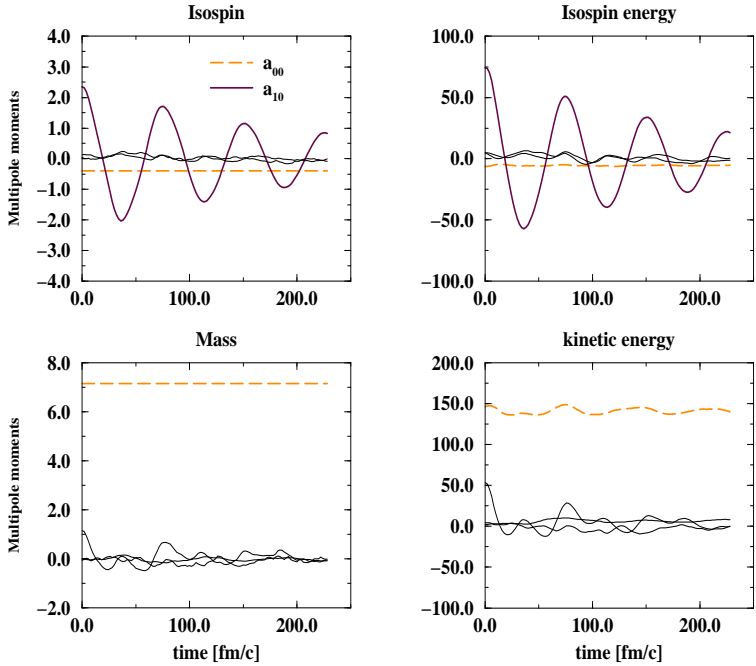,width=9cm,height=10cm,angle=-90}
}
\hspace{1.5cm}
\parbox[t]{5cm}{
\caption{The time evolution of multipole 
  moments for the isovector dipole excitation of $^{90}_{40}Zr$
  analogously to figure~\protect\ref{b212}. \label{b213}}
}}
\end{figure}
Like the monopole case we now investigate the mass number dependence.
With the same parameterization as for the monopoles, the experimental mass 
dependence of the energy is well reproduced, figure \ref{berman.3}. The damping is 
again under-predicted. The Vlasov simulation leads to a nearly constant width of 2 
MeV. This is again a hint that collisions cannot be neglected.
The isospin dependency of the dipole mode is plotted in
figure~\ref{dipoliso2} and is characterized by an increasing width and
decreasing centroid energy with increasing asymmetry.
\begin{figure}[h]
\parbox[]{18cm}{
\parbox[]{10cm}{
\psfig{file=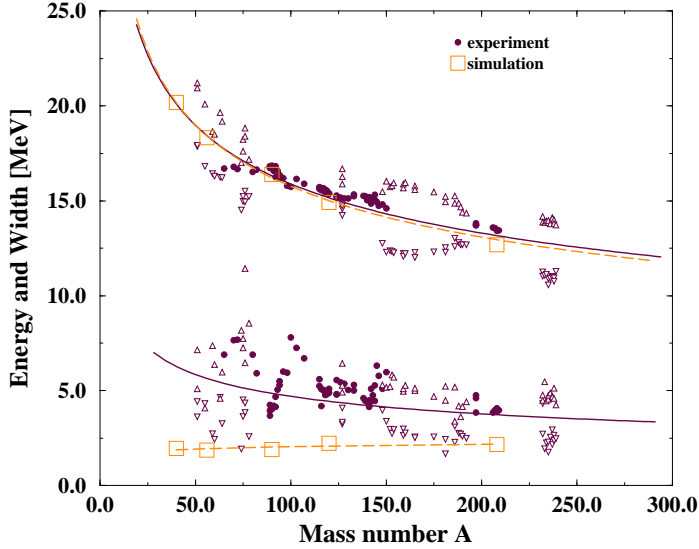,width=7.5cm,angle=-90}
}
\hspace{1.5cm}\parbox[t]{6cm}{
\caption{The experimental mass number dependence of isovector giant dipole 
oscillations (upper curve) is compared with the Vlasov simulation results.
The oscillations in the experimental damping width are due to shell closures.  
 \label{berman.3}} 
}}
\end{figure}

\begin{figure}[h]
\parbox[]{18.5cm}{
\parbox[]{6cm}{
\psfig{file=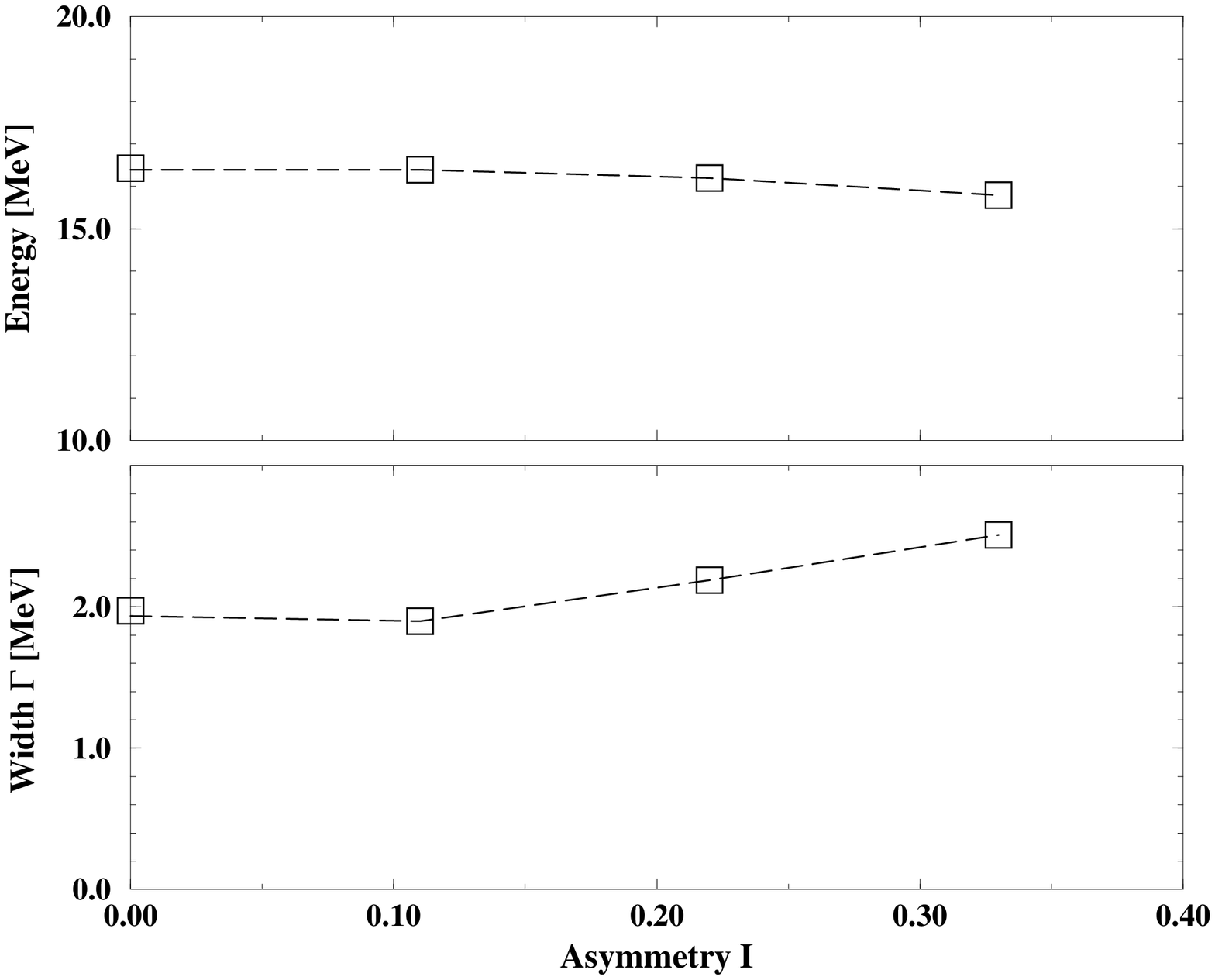,height=6cm,width=6cm,angle=-90}
}
\parbox[]{6cm}{
\psfig{file=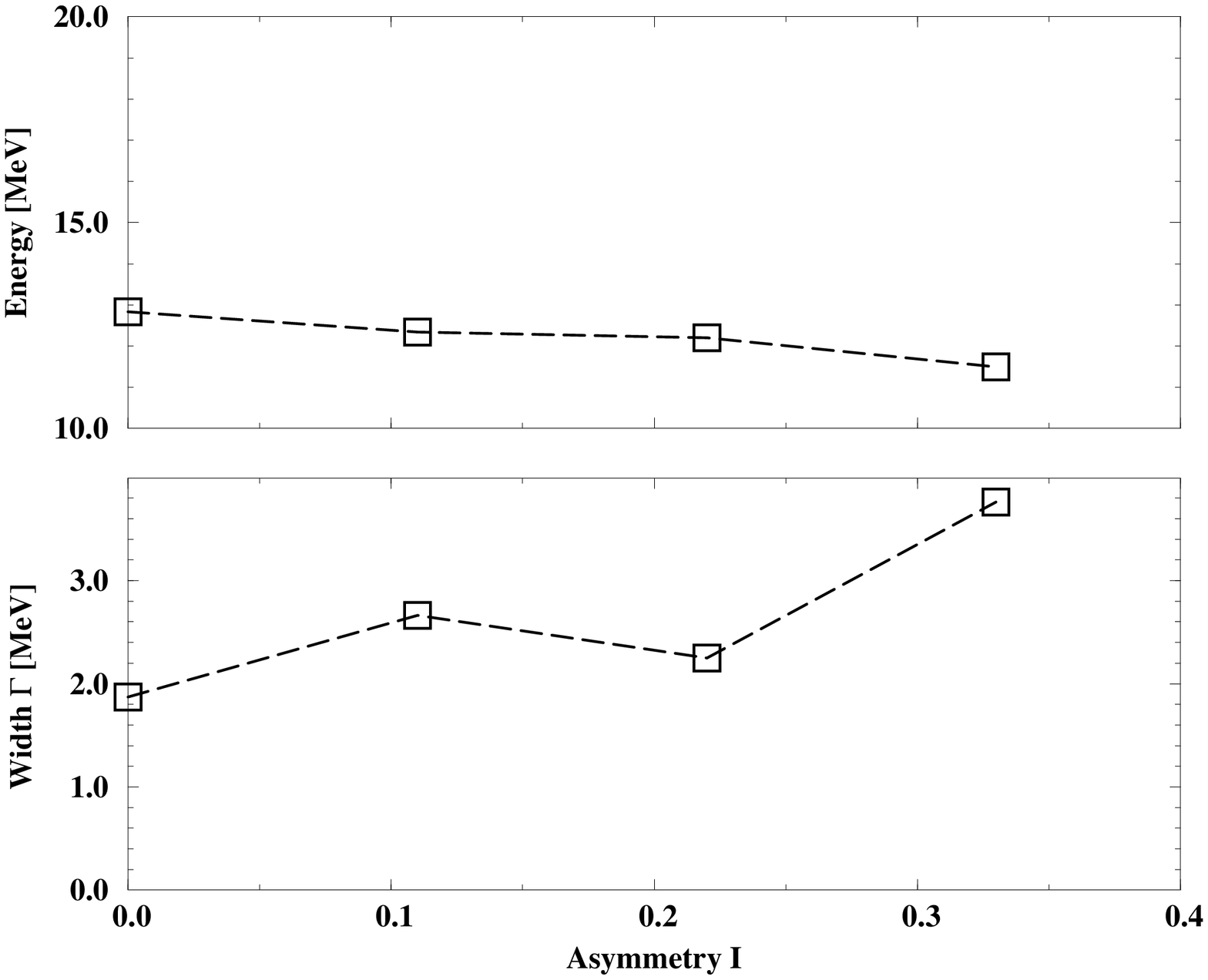,height=6cm,width=6cm,angle=-90}
}
\parbox[]{6cm}{
\psfig{file=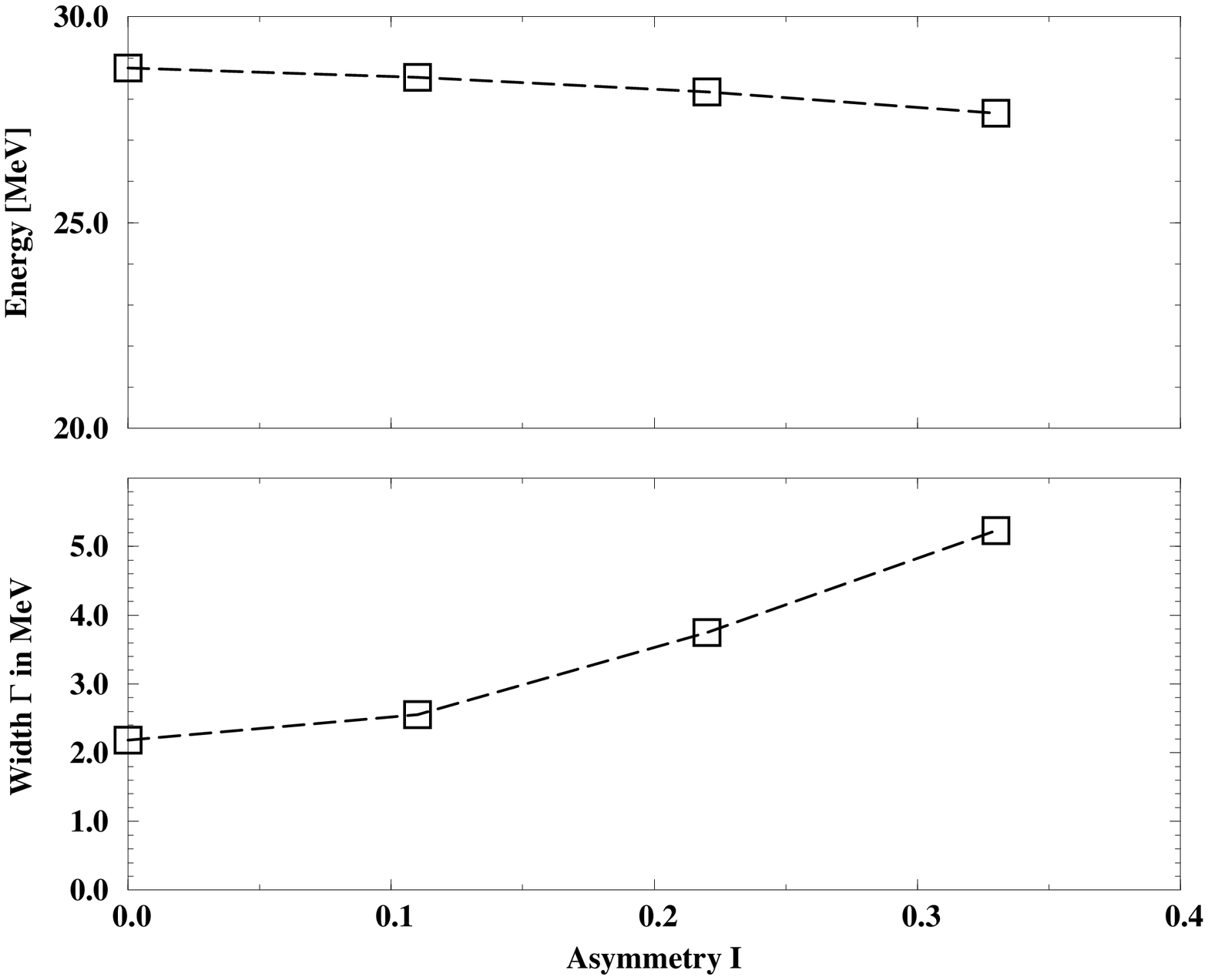,height=6cm,width=6cm,angle=-90}
}}
\caption{The isospin dependency of the centroid energy and the damping
  of the isovector dipole (left),  isoscalar quadrupole (middle) and
  isovector quadrupole mode (right). \label{dipoliso2}}
\end{figure}

\section{Quadrupole resonances}

Let us now investigate the quadrupole resonances. For these we distinguish
isoscalar and isovector modes. The quadrupole resonances are excited by
dividing the spatial distribution into two equal pieces which are
accelerated in opposite directions. In the first example the neutrons and protons are in
phase giving rise to an isoscalar mode.

\begin{figure}[h]
\parbox[]{18cm}{
\parbox[]{10cm}{
\psfig{file=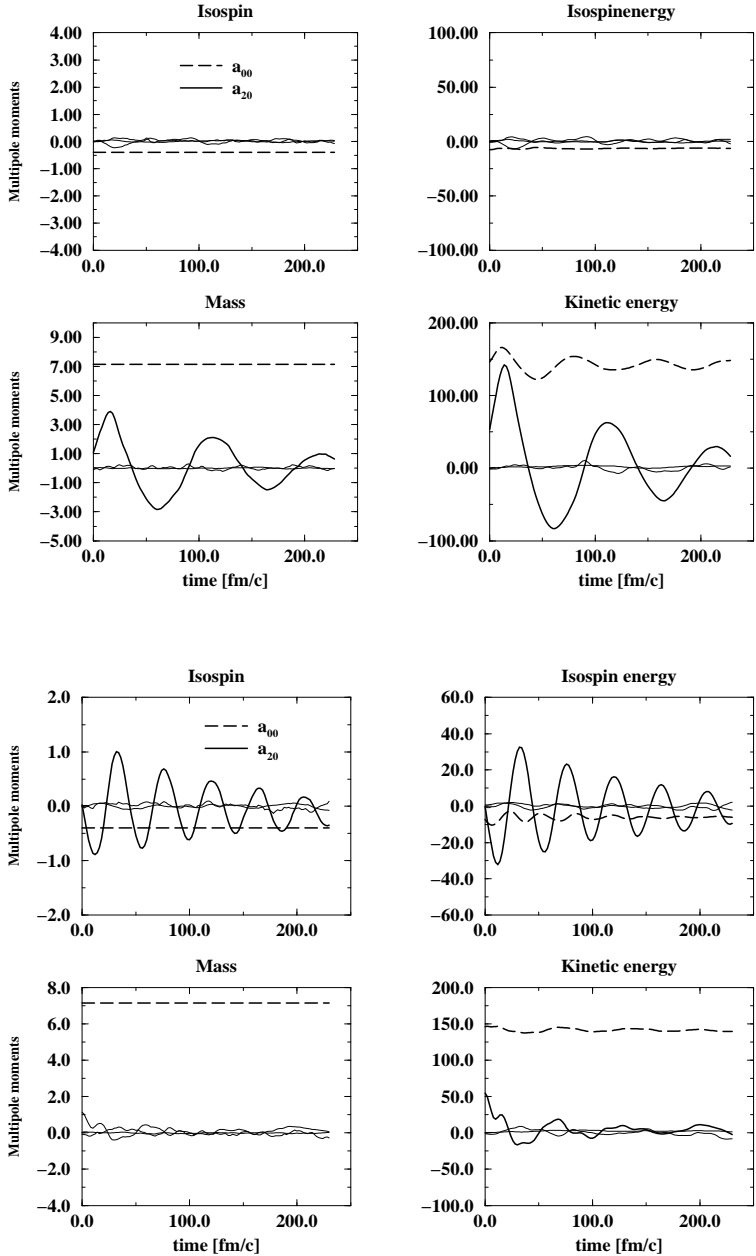,width=8cm,height=10cm,angle=-90}
}
\hspace{2.5cm}
\parbox[t]{5cm}{
\caption{The time evolution of multipole 
  moments for the isoscalar quadrupole excitation of $^{90}_{40}Zr$
  analogously to figure~\protect\ref{b212}. \label{b217}}
}}
\end{figure}

We see from the kinetic energy in figure~\ref{b217} that besides the
clear isoscalar quadrupole mode there is also a weak isoscalar
monopole mode excited.
The isospin dependency of the isoscalar quadrupole mode is plotted in
figure~\ref{dipoliso2}. As in the case of the isovector dipole
mode one sees a slight
decrease of the centroid energy and an increase of damping with
increasing asymmetry.

\begin{figure}[h]
\parbox[]{18cm}{
\parbox[]{10cm}{
\psfig{file=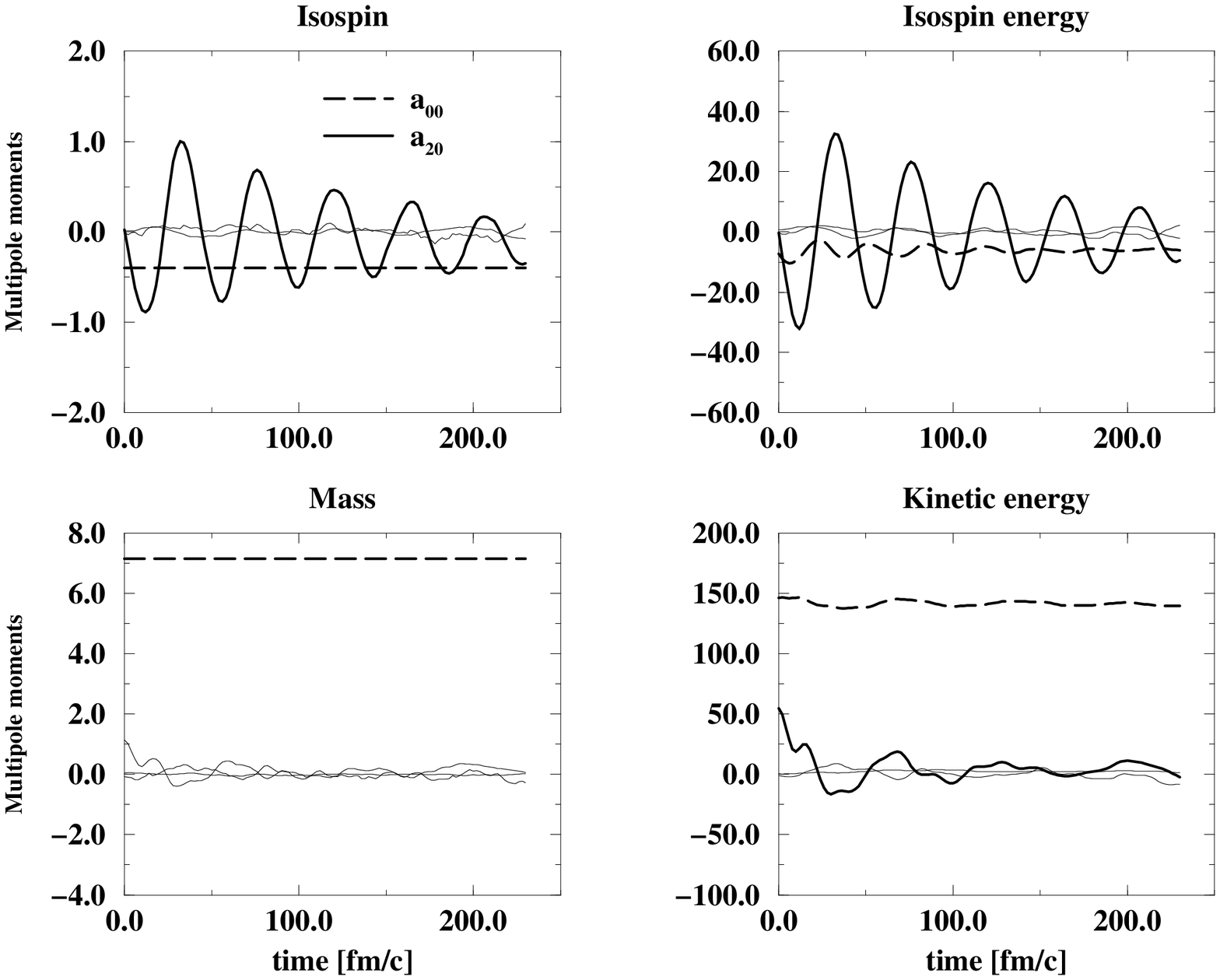,width=8cm,height=10cm,angle=-90}
}
\hspace{2.5cm}
\parbox[t]{5cm}{
\caption{The time evolution of multipole 
  moments for the isovector quadrupole excitation of $^{90}_{40}Zr$
  analogously to figure~\protect\ref{b212}. \label{b221}}
}}
\end{figure}

In the second case we excite protons and neutrons out of phase which
excites isovector quadrupole oscillations in figure~\ref{b221}. 
The isospin energy and the isospin show nice
oscillations in the quadrupole multipole moment. There is a slight
excitation of an isovector dipole mode too. We would like to remark that
due to the test particle width there occurs a mode coupling
\cite{MP96} which can
be seen in the slight excitation of isoscalar modes in the kinetic energy.
The isospin dependence shown in figure~\ref{dipoliso2} has the same
qualitative behaviour as the isoscalar mode, but is stronger in this case.

\chapter{Collective modes in linear response}

Now we would like to focus on small amplitude oscillations and we will develop
the linear response theory. Large amplitude motions and effects beyond
linear response are discussed in chapter \ref{oct}. 
The 
simplest microscopic theory which provides 
basic experimental features and which allows to include collisional
correlations is the Fermi-gas(liquid) 
model including dissipation. We will compare the results from this
linear response with the simulation of finite nuclei, and both will be
compared with
the experimental data.

The principle of linear response is easily explained. When a system
which is described by the kinetic equation (\ref{klassVlasov}) is
disturbed by an external potential $U^{\rm ext}$ it will react and will create a
density change ${\delta\rho}$. The connection of the latter to the
external perturbation is called the response
function, ${\delta \rho}=\chi U^{\rm ext}$. Without mean field in
(\ref{klassVlasov}) we would obtain the polarization $\Pi$ of the
system. The mean field introduces a selfconsistency such that the
relation between response function and polarization becomes
\be
\chi={\Pi \over 1- {\delta U \over \delta n}\Pi}
\label{cc1}
\ee
where we have assumed that the mean field $U$ is only density
dependent. Therefore the potential for the response function is
$V={\delta U \over \delta n}$. The collective mode of a system is now
characterized by the condition that the denominator in (\ref{cc1})
vanishes \be
\epsilon({\bf q},\omega)=1-V({\bf q}) \Pi({\bf q},\omega)=0
\label{dis}
\ee
since then an infinitesimal small external potential can
create a finite density oscillation due to the diverging response function.
Therefore we will search for complex zeros of the dielectric (or
dinuclic) function (DF),
$\epsilon({\bf q},\Omega+i \gamma)=0$, which provides us with the energy $\Omega$ and 
the damping
$\gamma$ of the collective mode.
The strength function is then given by
\be
S({\bf q},\omega)=\displaystyle{\frac 1 \pi {{\rm Im}\Pi
                  \over(1-V {\rm Re}\Pi)^2+(V{\rm Im}\Pi)^2}}.
       \label{strenght}
\ee
The equation of state like the isothermal compressibility can be expressed by
\be
&&\kappa={1\over n^2}\left ({\partial n\over \partial \mu}\right )_T={1\over n^2 
T}\lim\limits_{q\to0}\int {d\omega\over \pi} {1\over {\rm e}^{\omega/T}-1} {\rm 
Im}{\chi(q,\omega)}
\nonumber\\&&
\label{kap}
\ee
and the fluctuations and diffusion coefficients can also be
expressed via the response function.

Two lines of theoretical improvements of the response function can be
found in recent publications. The first one starts from TDHF equations and
considers the response of nuclear matter described by a non time-reversal
Skyrme interaction \cite{HNP96,GNN92,BVA95}. The other line tries to
improve the response by the inclusion of collisional correlations
\cite{HPR93,Mer70,KH94,RW98,MF99} and by considering multicomponent systems
\cite{MWF97,HAE78,HNP97,CTL98,Br99}. 
In \cite{MF99} both lines of improvements have been combined into one 
expression. We follow this line and derive the response function from 
a kinetic equation including mean field (Skyrme) and collisional correlations.
Therefore we consider interacting matter which can be described by an
energy functional ( mean field) ${\cal E}$ originally introduced by
Skyrme \cite{S56,S59} and the residual interaction. The latter we
condense in a collisional integral additional to the TDHF
equation. Then the response to an external perturbation will contain
the effect of Skyrme mean field and additionally the effect of the residual
interaction.

\section{Response function for asymmetric matter}

The system now consists of a number
of different species (neutrons, protons, etc...) interacting with their 
own kind and with the others. It is important to
consider the interaction between different
sorts of particles  if we want to include friction between different
streams of isospin components, etc. In particular the isospin current may
not be conserved in this way. Let us start with a set of coupled
quantum kinetic equations\footnote{
The quasiclassical Landau equation (\protect\ref{klassVlasov}) follows from gradient 
expansion as
\be
{\partial \over \partial t} f+\partial_{\bf p} \epsilon \partial_{\bf r} 
f-\partial_{\bf r} \epsilon \partial_{\bf p} f=-{f-f^{\rm l.e.}\over \tau}
\ee
} for the reduced density operator
$\hat \rho_a$ of a certain species denoted by the subscript $a$
\beq
\partial_t \hat \rho_a(t) =i[\hat \rho_a,{\hat E}_a(t)+{\hat U}_a^{\rm 
ext}]-\sum\limits_b\int\limits_0^t
dt'{\hat\rho_a(t')-\hat \rho_b^{\rm l.e.}(t') \over \tau_{ab}(t-t')}\label{kin}
\eeq
where ${\hat E}={\hat P}^2/2m+{\hat U}$ denotes the kinetic as well as
mean field energy
operator and the external
field which is assumed
to be a nonlinear function of the density. We approximate the
collision integral by a non-Markovian relaxation time which is derived in appendix 
\ref{ap}. This
turned out to be necessary to reproduce damping of zero sound
\cite{AB92,MTM96}. It accounts for the fact that during a two-particle
collision a collective mode can couple to the scattering process.
Consequently, the dynamical
relaxation time represents the physical content of a hidden three-particle 
process and is equivalent to memory effects. 

Furthermore, we assume
relaxation towards a local equilibrium 
\be
f_a^{\rm l.e.}({\bf p,R},t)=\sum\limits_q {\rm e}^{i{\bf q R}} \langle p+{q\over 
2}|\hat \rho_a^{\rm l.e.}|p-{q\over 2}\rangle=
f_0\left ({\varepsilon_a({\bf p}-{\bf Q}_a({\bf R},t))-\mu_a({\bf R},t)\over T_a({\bf 
R},t)}\right )
\label{2}
\ee
with the Fermi distribution $f_0$.
This local equilibrium will be specified here only by a small
deviation of the chemical potential of species $a$ ensuring density
conservation \cite{MF99}. The extension of the method including
further conservation laws and specifying also the local current and
the temperature can be found in \cite{MF99}.

Linearizing the kinetic 
equation (\ref{kin}) we obtain the matrix equation for the density
deviation $\delta \rho_b$ due to an external perturbation
$U^b_{\rm ext}$
\beq
\sum\limits_b \delta n_b \left\{
\delta_{ab}-{i \over \omega \tau_a+i} \left[\delta_{ab}
-{\tau_a \over \tau_b} C_{ab}\right] -\Pi_a(\omega+{i\over\tau_a}) 
                                               \alpha_{ab}\right\}
=\Pi_a(\omega+{i \over \tau_a}) U_{\rm ext}^a.\label{resp}
\eeq
The matrix $C_{ab}$ is given by
\beq
C_{ab}=\sum\limits_c \left\{ 1\over\tau \right \}_{ac}
{\Pi_c\Big[(\omega+{i\over \tau_a})\frac{m_a}{m_c}\Big]\over \Pi_c(0)}
\left\{ \frac{1}{\tau}\right\}^{-1}_{cb}\label{c1}
\eeq
with $\alpha_{ab}={\pa \rho_b} U_a$ the
linearization of the mean field with respect to the deviation of
density from equilibrium value caused by the external
perturbation $U_{\rm ext}$. The partial polarization function of 
species $a$ is
\beq
\Pi_a(\omega)=2 \intp\frac{f_a\ppl-f_a\pmi}{\e_a\ppl-\e_a\pmi-\omega+i0}.
\eeq
The factor 2 in front of the integral accounts for the spin
degeneracy according to $\rho_a=2 \intp f_a(p)$. 

Equation (\ref{resp}) represents the complete polarization of the
system, because $\delta \rho=\Pi \, U_{\rm ext}$. It represents a matrix
equation which is solved easily. The collective modes are given by
the zeros of the determinant of the matrix on the left hand side
of (\ref{resp}) because these are the poles of the polarization
matrix.
Since we
took into account relaxation processes between all species we are
able to cover current - current friction.
For a two component system, e.g. neutrons with density $\rho_n$ and
protons with density $\rho_p$, we write explicitly
\beq
&&(1-\Pi_n^{\rm M} \alpha_{nn})(1-\Pi_p^{\rm M}
\alpha_{pp})-(D_{np}+\Pi_n^{\rm M} \alpha_{np})(D_{pn}+\Pi_p^{\rm
M} \alpha_{pn})=0\label{dism}
\eeq
with the generalization of the Mermin polarization function \cite{M70} to
a multicomponent system
\beq
\Pi_a^{\rm M}={\Pi_a(\omega+{i\over \tau_a})\over
1-{i \over \omega\tau_a +i} (1- C_{aa})},
\label{mm}
\eeq
and the additional coupling due to asymmetry in the system
\beq
D_{np}={\tau_n\over \tau_p}{C_{np}\over C_{nn}-i \omega \tau_n}.
\label{D}
\eeq
The $D_{pn}$ are given by interchanging species indices.
This term does not appear for symmetric matter. 
Therefore we call this term the asymmetry coupling term further on.
The result (\ref{dism}) represents the generalization of the
dispersion relation (\ref{dis}) to a two - component system.

It is illustrative to recover known results for symmetric nuclear matter.
This is performed for the case of equal relaxation times
$\tau_p=\tau_n=\tau$ and equal deviation from the
mean field $\alpha_1=\alpha_{nn}=\alpha_{pp}$ and
$\alpha_2=\alpha_{np}=\alpha_{pn}$. Eq. (\ref{dism}) takes then the known Mermin form 
of dispersion relation \cite{M70}
\beq
1-(\alpha_{1}\pm\alpha_{2}) {\Pi(\omega+{i\over \tau})\over
1-{i \over \omega\tau +i} \left[1- {\Pi(\omega+{i\over
\tau}) \over \Pi(0)} \right]}=0\label{dis4}
\eeq
with the isovector mode $\alpha_{1}-\alpha_{2}$ and the
isoscalar mode $\alpha_{1}+\alpha_{2}$. Please note that for zero
temperature the Mermin expression (\ref{dis4}) agrees with the
response function derived from taking into account multipole expansion
in momentum of the disturbed distribution \cite{TKL99}.

We have presented a general dispersion relation for the
multicomponent system including known special cases. The
dispersion relation (\ref{dism}) is similar to the one derived
recently in \cite{CTL98,BAR98} if we neglect the collisional coupling
$D_{np}$. Also a more general polarization function (\ref{mm}) 
is presented here including collisions within a conserving approximation
\cite{HPR93}. In the
following we will apply this expression to the damping of giant
dipole resonances in symmetric as well as asymmetric nuclear
matter.

\section{Application to nuclear matter}

Before we can solve the dispersion relation (\ref{dism}) we have to
specify the wave vector characteristic of the considered mode.
This is performed according to the Steinwedel-Jensen \cite{SJ68} model. 
We assume that the surface remains constant and the density oscillation obeys a wave 
equation 
with the boundary condition that the radial velocity vanishes 
on the spherical surface with radius $R=1.13 A^{1/3}$. This leads to 
$j_l'(k R)\equiv 0$ with the spherical Bessel function of order 
$l=0,1...$ associated with the monopole, dipole... resonances. From this condition one obtains a connection between the wave vector
and the radius of the nuclei or the mass number in the form
 \be
k_{i,l}={c_{il}\over 1.13 A^{1/3}}
\ee
for the $i$-th mode of multipolarity $l$ where $c_{il}$ is the $i$-th
zero of $j_l'(c)\equiv 0$.

For the  monopole modes as compression modes we have no zero of first order 
for $l=0$.
For the dipole modes one has in first order $k=2.08/R$ which describes the 
giant isovector dipole resonance (IVGDR) while the ISGDR is a spurious mode 
in first order. This would just mean an unphysical oscillation of center of 
mass motion. However, in second order $k=5.94/R$ the isoscalar giant dipole 
resonance (ISGDR) has been observed recently (\cite{D97} and references therein). 
This can be considered as a density oscillation inside a sphere as we
will discuss in chapter \ref{isg}.

Besides the occurring wave vector we have also to specify the
relaxation times which contain the effect of collisions.
The dynamic relaxation time has been derived 
by Sommerfeld expansion \cite{FMW98} in appendix \ref{ap}  as
\beq
{1\over \tau_{ab}^{\rm gas}(\omega)}&=&{1\over \tau_{ab}(0)} \left[1+\frac 3 4
\left ({\omega \over \pi T}\right )^2 \right]
\qquad \qquad {1\over \tau_{ab}^{\rm liq}(\omega)}={2\over \tau_{ab}(0)} \left[1+\frac 
1 2 
\left ({\omega \over \pi T}\right )^2 \right]\label{mem}
\eeq
for $a,b$ neutrons or protons respectively and dependent on whether we use a Fermi liquid 
or Fermi gas model. The Markovian relaxation time 
is given in terms of the cross section $\sigma_{ab}$ between species $a$ 
and $b$ as $\tau^{-1}_{ab}={4 m\over 3 \hbar^3}\sigma_{ab} T^2$.

We will use as an illustrative example a slightly different mean field
potential
than (\ref{hf}) given by \cite{VB72,BV94}
\beq
U_a=t_0 \left [ (1+{x_0\over 2}) (\rho_n+\rho_p)-(x_0+\frac 1 2) \rho_a \right ]
+{t_3 \over 4} \left[(\rho_n+\rho_p)^2-\rho_a^2\right]
\label{vaut}
\eeq
with $a=n,p$ the density of neutrons or protons respectively.
The Coulomb interaction leads to an additional contribution for
the proton mean-field
\beq
U_p^{C}(q)={4 \pi e^2\over q^2} \rho_p(q).
\label{co1}
\eeq
The model parameters used here reproduce the Weizs\"acker formula
\beq
{{\cal E}\over A}= -a_1 +{a_2\over A^{1/3}}+{a_3 Z^2\over A^{4/3}} + a_4 I^2
\eeq
with the volume energy $a_1=15.68$ MeV, Coulomb energy $a_3=0.717$ MeV and 
the symmetry energy $a_4=28.1$ MeV with the asymmetry parameter
$I={n_n-n_p\over n_n+n_p}$.

\begin{figure}[h]
\parbox[]{18cm}{
\parbox[]{10cm}{
\psfig{file=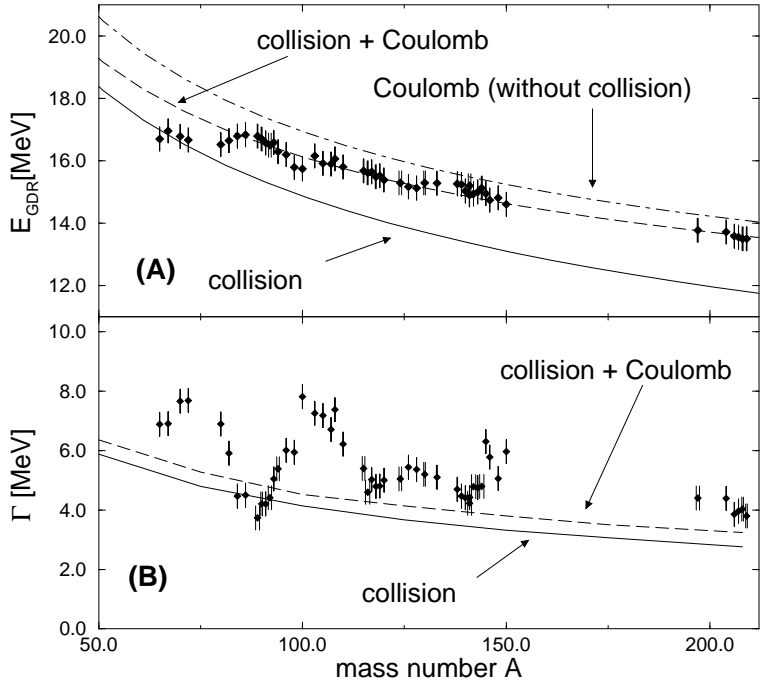,height=10cm,angle=-90}
}
\hspace{2.5cm}
\parbox[t]{5cm}{
     \caption{The experimental centroid energies {\bf (A)} 
              and damping rates {\bf (B)} [filled symbols] of the giant dipole
              resonances vs mass number (data from Ref. \protect\cite{BER88})
              together with different approximations at $T=0$. }
             \label{energie}    }
}
\end{figure}

The energy and damping rates are now determined by the zeros of the
(Mermin) polarization function (\ref{dis4}). First we plot the solution of the 
dispersion relation
(\ref{dism}) for symmetric nuclear matter. 
In figure \ref{energie} 
we have plotted the real and imaginary (FWHM) 
part of complex energy for different approximations with relaxation time 
(collisions) and with and without Coulomb mean field (\ref{co1}). In figure \ref{energie} {\bf (A)} we 
find that the
inclusion of Coulomb effects reproduces the experimental shape of 
the centroid energies at higher mass numbers (dot-dashed line). Taking
only collisions into account fails to describe higher mass numbers 
(solid line).
Considering Coulomb together with collisions (dashed line), the centroid
energies are  reduced towards the data.   

The experimental values of the damping
rates are also presented versus mass number (figure \ref{energie} {\bf (B)}). 
We recognize that the FWHM is just twice the damping rate $\Gamma=2
\gamma$
which has been recently emphasized \cite{TKL97}.
It has to be stressed that the experimental data are accessible by this FWHM.
Here we have considered only collisions and have a vanishing Landau damping for
the infinite matter model \cite{FMW98}.

\section{Comparison between simulation and linear response}

We have already stressed that the simulation of finite nuclei
neglecting collisional correlations lead to a damping of about
$\Gamma_{\rm Land}=2$MeV
for the giant dipole resonance at zero temperature. This was
attributed to the surface effects of spherical symmetric nuclei. We
consider this as the numerical value of the contribution attributed to
the wall \cite{TKL99} at zero temperature. For finite temperature the
will be considered an additional contribution due to thermally
deformed shapes
in section \ref{exit1}.
Nevertheless it is now interesting to compare the isospin dependences
of the simulation without collisions and the linear response
including collisions but no surface effects.

\begin{figure}[h]
\parbox[]{18cm}{
\parbox[]{10cm}{
\centerline{\psfig{file=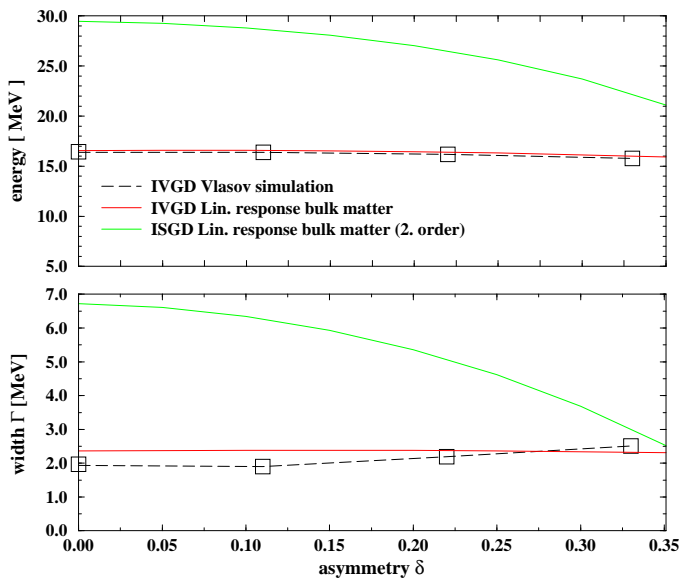,height=9cm,angle=-90}}
}
\hspace{2.5cm}
\parbox[t]{5cm}{
     \caption{The dependence of isovector dipole resonance energy and
       damping on the asymmetry obtained from simulations and from
       linear response for $A=90$.\label{d1}}
}}
\end{figure}

In figure \ref{d1} we see that the damping is slightly increasing with
increasing asymmetry for simulations without collisions while the
collisional contributions cause a slight decrease of damping. This
decrease is much more pronounced in the second order isoscalar dipole
mode.
The energy in turn shows the same behaviour for simulations without
correlations and for linear response with correlations; it slightly
decreases the value.

\begin{figure}[h]
\parbox[]{18cm}{
\parbox[]{10cm}{
\centerline{\psfig{file=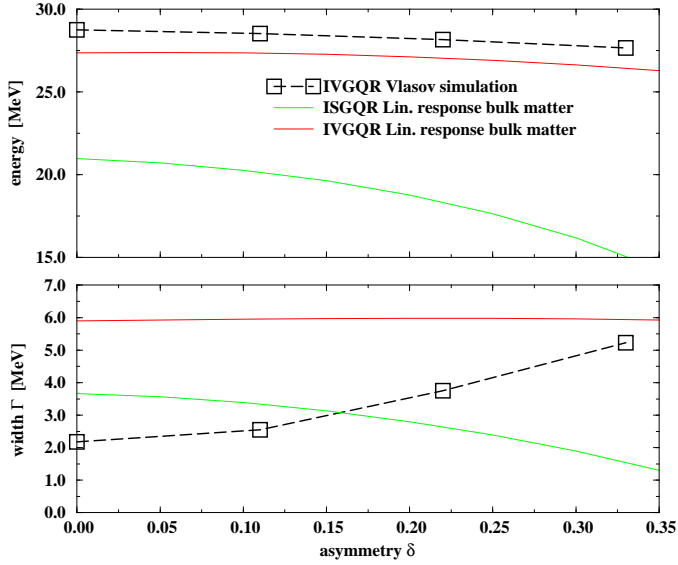,height=9cm,angle=-90}}
}
\hspace{2.cm}
\parbox[t]{5.5cm}{
     \caption{The dependence of isovector and isoscalar quadrupole resonance energy and
       damping on the asymmetry obtained from simulations and from
       linear response for $A=90$.\label{d2}}
}}
\end{figure}

The same qualitative feature can be recognized for the quadrupole
modes in figure (\ref{d2}). However the importance of collisional
damping is in this case much larger than the finite size effects.

\section{Giant resonances in excited nuclei}

\begin{figure}[h]
\parbox[]{18cm}{
\parbox[]{9cm}{
\centerline{\psfig{file=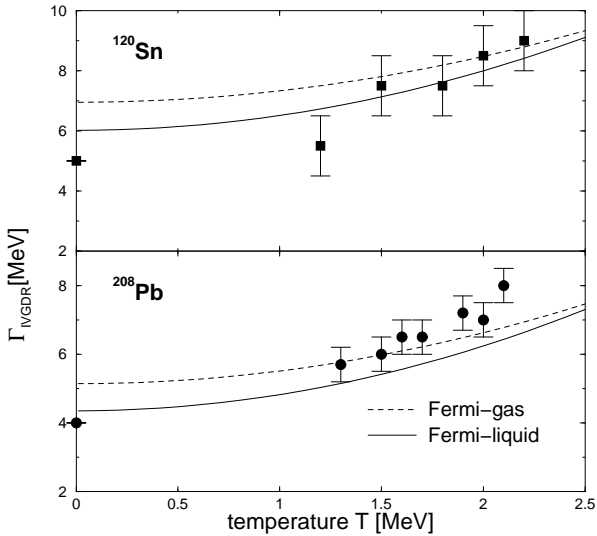,height=8cm,angle=-90}}
}
\hspace{0.5cm}
\parbox[t]{8cm}{
 \caption{Experimental damping rates  vs. temperature of IVGDR 
          for $^{120}$Sn and $^{208}$Pb  
          ($^{120}$Sn from Ref. \protect\cite{RAP96} and $^{208}$Pb from
          Ref. \protect\cite{RAM96}) compared with the solution of the dispersion
          (\protect\ref{dis4})
          relation $\Gamma=2\,\gamma$  for the Fermi gas (dashed lines) and
          the Fermi liquid model (solid lines). Please note that this
          damping is reduced when the momentum conservation is
          included like it is done in figure \protect\ref{pbt}.\label{bild1}}                      
}}
\end{figure}

We consider now the temperature dependence of the GDR. We have found
that the centroid energy only slightly decreases with increasing energy.
In Fig. \ref{bild1} the theoretical 
damping rates $\Gamma=2\,\gamma$
of the IVGDR modes in $^{120}$Sn and 
$^{208}$Pb are plotted as a function of temperature together with experimental values.
The results of the Fermi gas model (\ref{avt}) and 
the Fermi liquid model (\ref{avtliq}) are very close and in good
agreement with the data at low temperature. The higher temperature dependence still remains too flat 
compared to the experiments. The small difference between both models for T=0 
vanishes with increasing temperature.

While the figures indicate that the linear response including
collisional correlations together with the Steinwedel Jensen model can
describe quite well the gross features of giant dipole resonances we
should keep in mind that we have neglected here the surface
effects. From the simulation results of the second chapter we have seen
that there is a damping of about $\Gamma_{\rm Land}=2$MeV for finite
nuclei at zero temperature without
collisional correlations. Simply adding now both contributions would
overestimate the experimental damping. We would like to point out that
the inclusion of only density conservation in the derivation of the
response function so far is overestimating the width. In \cite{MF99} we have
shown that the inclusion of momentum conservation reduces this width
by approximately $2$MeV. Therefore we might anticipate that both
the finite size effects together with the collisional
contribution lead to the correct value for low temperatures.
\begin{figure}[h]
\parbox[]{18cm}{
\parbox[]{10cm}{
\psfig{file=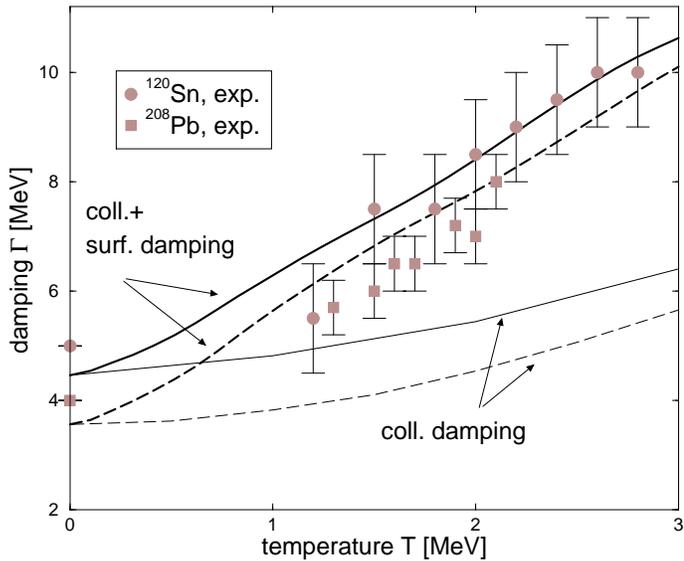,height=9cm,angle=-90}}
\hspace{1.5cm}
\parbox[t]{6cm}{
     \caption{The effective damping consisting of only collisional damping 
              compared with collisional and surface 
              damping [$^{120}$Sn (solid lines) and $^{208}$Pb (dashed lines)]
              together with the experimental data (filled symbols: Sn 
              from Ref. \protect\cite{RAP96} and Pb from 
              Ref. \protect\cite{RAM96}). } \label{pbt} }
}   
\end{figure}
In figure
\ref{pbt} we see that the temperature increase is too flat compared to
the experimental finding if we consider only collisional damping (thin
lines). From simulation results of ground state we have
already found a contribution from finite size effects. We might now
expect that the discrepancy in temperature behaviour is due to finite
size effects. This should be explained and understood within a more simplified 
model.

\section{Nuclear surface contribution - excited nuclei}\label{exit1}

Besides the $\Gamma_{\rm Land}=2$MeV of the simulation in the ground
state we can have additional damping due to shape deformation if we
increase the temperature. In  Ref. \cite{MVFLS98} we have presented a method to include also 
scattering with the non-spherical nuclear surface. This improves the temperature dependence 
remarkably (thick lines) in figure 
\ref{pbt}.

The idea is to consider the surface scattering contribution
to the damping determined by the additional Lyapunov exponent due to the surface. It 
has been shown that
such Lyapunov exponents appear in the response function additive to
the frequency as imaginary shift provided the Lyapunov exponent is
small compared to the product of wave vector times Fermi velocity \cite{MVFLS98,M99}
\be
\Pi_0^{\rm surf}(q,R,\omega)=\Pi_0^{\rm inf}(q,p_f(R),\omega+i \lambda)
\label{ldac}.
\ee
The Lyapunov exponent by itself is calculated for different deformations of the 
surface corresponding to different temperatures
\be
R_{\lambda}(\theta)
=R_0\,\left(1+\alpha_{00}+
             \alpha_{\lambda}\,P_{\lambda}(\cos{(\theta)})\right)\label{r1}
\ee
with the nuclear radius $R_0=1.13 A^{1/3}$fm, and where
$\lambda=2$ corresponds to the quadrupole and 
$\lambda=3$ to the octupole deformation \cite{ringschuck}.
The corresponding mean deformation \footnote{For small deviations we found identical 
Lyapunov exponents for prolate $\alpha>0$ and oblate $\alpha<0$ deformations and 
therefore we do not distinguish the sign of $\alpha$.} is linked to the temperature
\be
\langle\alpha\rangle={\int d \alpha |\alpha| \exp{(-E_B(\alpha)/T)} \over 
\int d \alpha \exp{(-E_B(\alpha)/T)}}
\label{38}
\ee
where the surface dependent energy $E_B(\alpha)$ is given by the Bethe-Weizs\"acker 
formula \footnote{Please remember that in principle the Coulomb energy changes with 
small deformation as well according to the factor \protect\cite{greinermaruhn}
$
1-5 {(\lambda-1)/ (2 \lambda+1)^2} \, \alpha_{\lambda}^2
$
while the surface term changes as 
$
1+(\lambda-1)(\lambda+2)/2/(2\lambda+1) \alpha_{\lambda}^2.
$
Only the latter correction is considered since the Coulomb energy deformation would lead to 
corrections of around $0.3\%$ and are neglected here.}
\be
&&
{E_B}(\alpha)= -a_1 +{a_2\over A^{1/3}}+{a_3 Z^2\over A^{4/3}} + a_4 
I^2+ a_5 A^{2/3} {S(\alpha) \over S(0)}\nonumber\\&&
\ee
with the volume energy $a_1=15.68$ MeV, Coulomb energy $a_3=0.717$ MeV, 
the symmetry energy $a_4=28.1$ MeV and the surface 
energy $a_5=18.56$ MeV.
For the quadrupole, $S_2$, and octupole, $S_3$, deformations one gets
\be
{S_2(\alpha) \over S(0)}&=&1+\frac 2 5 \alpha^2+
                                                   {\cal O}(\alpha^3)\nonumber\\
{S_3(\alpha) \over S(0)}&=&1+\frac 5 7 \alpha^2+
                                                   {\cal O}(\alpha^3). \label{cs}
\ee
This represents the lowest order expansion in $\alpha$, 
however the next term gives already corrections in fractions of percent
for the highest deformations considered here.
By this way, the statistical model (\ref{38}) leads to a connection between 
temperature and 
deformation as
\be
T=c \pi a_5 A^{2/3} <\alpha>^2\label{T}
\ee
where the constant $c$ is given by the coefficient of $\alpha^2$
in Eq. (\ref{cs}) for the corresponding quadrupole or octupole deformation.

\begin{figure}[h]
\parbox[]{18cm}{
\parbox[]{10cm}{
\psfig{file=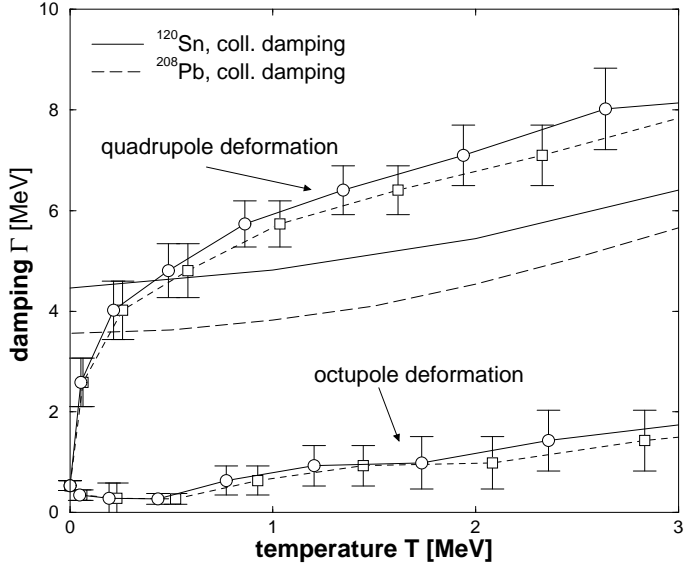,height=9cm,angle=-90}}
\hspace{1.5cm}\parbox[t]{6cm}{
  \caption{The scaled collisional damping is
           compared with the damping according to the 
           chaotic scattering from the surface of quadrupole and 
           octupole deformed shapes for $^{120}$Sn and $^{208}$Pb.\label{fig3}}}} 
\end{figure}

Using Eq. (\ref{T}) we can translate 
the Lyapunov exponent $\lambda$ calculated as a function of deformation into a 
function of the temperature. 
In figure~\ref{fig3} the contribution to 
the damping of IVGDR for $^{120}$Sn (circles) and $^{208}$Pb (squares) 
is presented for different shape deformations versus temperature.
If we add quadrupole and octupole deformations 
we come up with a damping curve very 
similar to \cite{OBB97}. The damping starts at zero and 
increases rapidly with increasing temperature. We see 
that the main contribution comes from the quadrupole deformation while 
the octupole deformation is only sizeable at higher 
temperature. Let us note that the qualitative difference between 
Sn and Pb is reproduced by including surface scattering as well as collisional damping.

The collisional contribution $\Gamma=2 \gamma$ is
plotted as well (Sn: solid line, Pb: dashed line). We recognize that
both contributions by themselves, collisional as well as surface scattering, account 
almost for the same amount required by the experimental values, see figure \ref{fig3}. 
A proper relative weight between both processes is therefore necessary which will be 
introduced in the following.

Let us note before that we have considered the surface contribution to
the damping for a deformed surface by temperature which therefore
vanishes at zero temperature. However, for a
spherical surface we got already from simulation a zero temperature
contribution of $\Gamma_{\rm Land}=2$MeV. Consequently for the total damping
caused by surface both contribute. 

So far we have not considered that only particles close
to the surface can appreciably contribute to the
surface scattering and to this chaotization process, while particles deep inside
the nuclei are screened out of this process.
Consequently we consider the corresponding collision frequencies as the measure to 
compare surface collisions with inter-particle collisions. The collision frequency 
between particles is given by $1/\tau_0$ of (\ref{mem}). The collision frequency of 
particles with the deformed surface beyond a sphere, $\nu_{\rm surf}$, is given 
by the product of the density with the surface increase
$S(\alpha)-S(0)=c \alpha^2 4 \pi R_0^2$ according to Eq. (\ref{cs}) and with
the mean velocity in radial direction $v_r=3/8 v_F$. The result is
\be
\nu_{\rm surf}=1.5 T n_0 v_F r_0^2/a_5
\label{surf}
\ee
where we have used Eq. (\ref{T}) to replace $\alpha$.
We see that the frequency (\ref{surf}) is independent of the size of the nucleus and 
linearly dependent on the temperature.
We use the ratio of these two frequencies to weight properly the two
damping mechanisms, the surface collisional, $\lambda$, and
the inter-particle collisional, $\gamma$, contributions. Consequently
the full - width - half - maximum (FWHM) reads 
\be
\Gamma_{\rm FWHM}&=&2 \zeta \, \gamma+2 (1-\zeta) \,
  (\lambda)+2 \gamma_{\rm Land} \nonumber\\
&\equiv&\Gamma_{\rm coll}+\Gamma_{\rm surf}+\Gamma_{\rm Land}.
\label{eff}
\ee
With the help of (\ref{mem}) and (\ref{surf}) the weighting factor $\zeta$ is given by
\be
 &&\zeta(T)={{1\over \tau_0(T)}\over {1\over \tau_0(T)}+\nu_{\rm surf}(T)}.
\ee
One sees that for zero and high temperatures $\zeta=1$ and 
due to Eq. (\ref{eff}) only the collisional contributions matter. 
Since $\nu_{\rm surf}$ is linear in the temperature and $1/\tau$ depends quadratically 
on the temperature, the weighting factor $\zeta$ has a minimum at temperatures around 
$T_c={\sqrt{3} \over 2 \pi}\omega $ for the gas model (\ref{mem}) and the surface contributions become important. 
In the case of the IVGDR this corresponds to a temperature of $T\approx 3.7$MeV, which 
is the upper limit of currently achievable experimental temperatures. Therefore we can 
state that at low and high temperatures the collisional damping is dominant while for 
temperatures around $T_c$ the surface contribution becomes significant.

\begin{figure}[h]
\parbox[]{18cm}{
\parbox[]{12cm}{
\centerline{\psfig{file=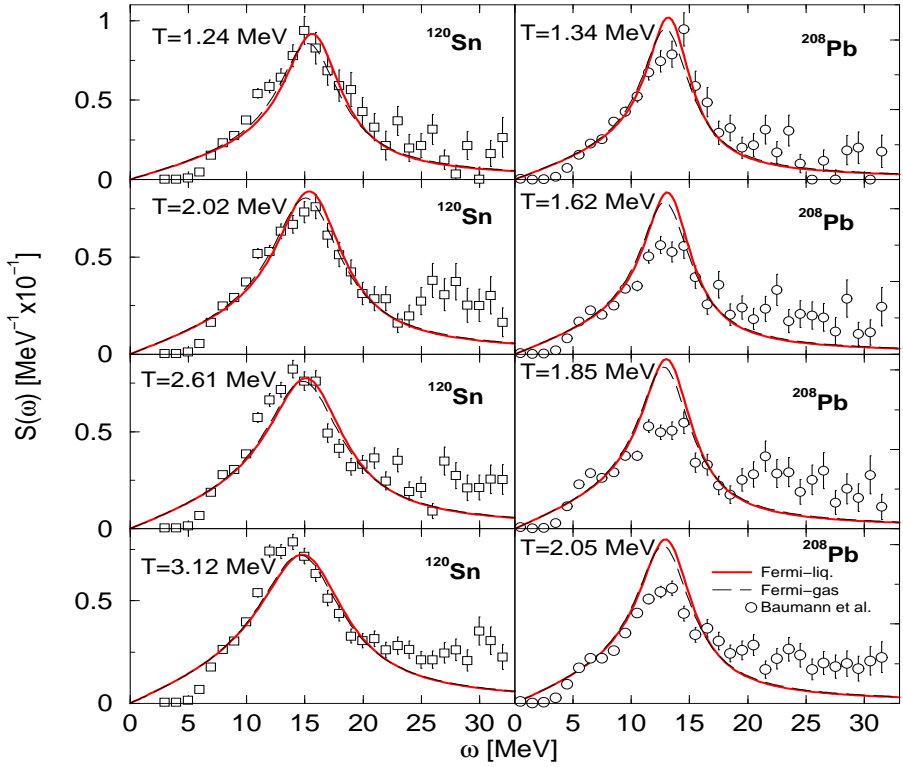,height=10cm,width=12cm}}}
\hspace{0.5cm}
\parbox[t]{5cm}{
 \caption{The IVGDR strength function in  $^{120}$Sn (LHS) and $^{208}$Pb (RHS)
          within the Fermi gas model (dashed lines) and Fermi liquid model 
         (solid lines) at several temperatures compared with normalized
          data from Ref. \protect\cite{BAU98}. \label{bild2}}}}
\end{figure}

In figure \ref{pbt} we have compared the effective damping  
according to Eq. (\ref{eff})
with the experimental data. Let us recall that the collisional damping
value is reduced here since we include momentum conservation too for
the Mermin polarization function. On the other side the surface
damping consists now of the deformation contribution by temperature
and by the zero temperature contribution obtained by simulation of Vlasov
equation. We find a reasonable quantitative agreement combining all
these parts. 
This is illustrated more in detail in Fig. \ref{bild2} where we have plotted the strength function (\ref{strenght})
for $^{120}$Sn (LHS) and $^{208}$Pb (RHS)
within the Fermi gas model (\ref{avt}) (dashed lines) and 
Fermi liquid model (\ref{avtliq}) (solid lines) with the normalized data from
Ref. \cite{BAU98}.
The good overall agreement of the shape evolution with the
experiment is again accompanied  by only minor differences between the
Fermi gas and the Fermi liquid model.

\section{Higher order modes - surface modes}\label{isg}

The existence of isoscalar giant dipole resonance (ISGDR) in nuclear
matter is considered as a spurious mode in most text books since
one associates with it a center of mass motion. The more surprising
was the experimental justification of a giant resonance carrying
the quantum numbers of a isoscalar and dipole mode \cite{HD81,D97,G99}. 
Consequently
one has to consider higher harmonics as explanation of such a
mode \cite{HD81,D73,HSZ98}. 
Usually this mode is associated with a squeezing mode analogous 
to a sound wave \cite{HD81,D97,GSA81,G99}.

In this chapter we want to discuss the influence of surface effects on
the ISGDR compression mode. We will show that even in the frame 
of the Fermi liquid model such modes can be understood. 
Moreover we claim that the surface
effects are not negligible for reproducing the strength function.
While we have already given a phenomenological approach to include
surface effects we want to show now a straightforward possibility to include
surface effects in the response function.
Consequently we give first a short derivation of response
function including surface effects. This will result in a new
formula in the temperature-dependent extended Thomas-Fermi 
approximation \cite{RS80}.

The starting point is again the semi-classical Vlasov equation
(\ref{klassVlasov}). Instead of using now a spatial homogeneous
equilibrium as done so far we consider now explicitly the spatial dependence.
Provided we know the response
to the external potential without self-consistent mean field, which
is described by the polarization function $\Pi$
\be
\delta n({\bf x},t)=\int d{\bf x}' \Pi({\bf x,x'},t) U^{\rm
ext}({\bf x'},t),
\label{po}
\ee
the response including mean field, $\chi$, is given by
\be
&&\chi({\bf x,x'},t)=\Pi({\bf x,x'},t)+\int d{\bf x}_1d{\bf x}_2
\Pi({\bf x},{\bf x}_1,t) {\delta
U^{\rm ind}({\bf x}_1,t)\over \delta n({\bf x}_2,t)} \chi({\bf
x}_2,{\bf x'},t).
\label{chi}
\ee

Therefore we concentrate first on the calculation of the
polarization function $\Pi$ and linearize the
Vlasov equation (\ref{klassVlasov}) according to $f({\bf
p,R},t)=f_0({\bf p,R})+\delta f({\bf p,R},t)$ such that the
induced density variation $\delta \rho({\bf R},t)=\int d{\bf p}
\delta f/(2 \pi \hbar)^3$ reads
\be
&&\delta \rho({\bf x},\omega)=\int{d{\bf q}\over (2\pi\hbar)^3} {\rm e}^{i{\bf
qx}}\int{d{\bf p}d{\bf x'}\over (2\pi\hbar)^6}{\rm e}^{-i{\bf qx'}}
{{\bf \nabla}_p
f_0({\bf p,x'}) {\bf \nabla}_{x'} U^{\rm ext}({\bf x'},t)
\over i(\omega-{{\bf pq}\over m})}.\nonumber\\
&&
\label{1}
\ee
Here we have employed the Fourier transform of space and time coordinates
of (\ref{klassVlasov}) to solve for $\delta f$ and inverse
transform the momentum into the form (\ref{1}).
Comparing (\ref{1}) with the definition
of the polarization function (\ref{po}) we extract with one
partial integration
\be
\Pi({\bf x,x'},\omega)=-{\bf \nabla}_{x'}\int{d{\bf p}d{\bf q}\over (2\pi\hbar)^6}
{\rm e}^{i{\bf q} ({\bf x-x'})}{{\bf \nabla}_p f_0({\bf p,x'})
\over i (\omega-{{\bf pq}\over m})}.
\label{p}
\ee
With (\ref{p}) and (\ref{chi}) we have given the polarization and
response functions for a finite system.

In the following we are interested in the gradient expansion
since we believe that the first order gradient terms will bear
the information about surface effects. Therefore we change to
the center of mass and difference coordinates ${\bf R}=({\bf
x}_1+{\bf x}_2)/2$,
${\bf r}={\bf x}_1-{\bf x}_2$ and retaining only first order
gradients we get from (\ref{p}) after Fourier transform of ${\bf r}$
into ${\bf q}$
\be
&&\Pi({\bf R,q})=-\int{d{\bf p}\over (2\pi\hbar)^3}{{\bf q \nabla}_p f_0({\bf
p,R})\over \omega-{{\bf p q}\over m}}
+{i\over 2} {\bf \nabla}_R
\int{d{\bf p}\over (2\pi\hbar)^3} \biggl ( {{\bf \nabla}_p f_0({\bf p,R})\over
\omega-{{\bf pq}\over m}}-{{\bf p}\over m} {{\bf q.\nabla}_p
f_0({\bf p,R})\over (\omega-{{\bf pq}\over m})^2} \biggr ) \nonumber\\
&=&-\int{d{\bf p}\over (2\pi\hbar)^3} [{\bf q}-i{\bf \nabla}_R (1+{\omega\over
2}\partial_{\omega})]{{\bf \nabla}_p f_0({p^2\over 2 m},{\bf R})\over
\omega-{{\bf pq}\over m}}
\label{grad}
\ee
where in the last equality we have assumed radial momentum
dependence of the distribution function $f_0$. We recognize that
besides the usual Lindhard polarization function as the first
part of (\ref{grad}) we obtain a second part which is expressed
by a gradient in space. The first part corresponds to the Thomas-Fermi 
result where we have to use the spatial dependence in the
distribution functions and the second part represents the
extended Thomas-Fermi approximation. So far we did not assume any
special form of the distribution function. Therefore the
expression (\ref{grad}) is also valid for any high temperature
polarization of finite systems.

What remains to be shown is that the response function (\ref{chi})
does not contain additional gradients. This is easily confirmed
by two equivalent formulations of (\ref{chi}),  $\Pi^{-1}\chi=1+V\chi$ and
$\chi\Pi^{-1}=1+\chi V$, which by adding yield the anticommutator
\be
[\Pi^{-1},\chi]_+=2+[V,\chi]_+.
\ee
This anticommutator does not contain any gradients up to second order. 
Therefore
we have [$V={\delta U^{\rm ind}/\delta n}$]
\be
 \chi({\bf R,q},\omega)={\Pi({\bf R,q},\omega)\over 1-V({\bf R,q},\omega)
\Pi({\bf R,q},\omega)} +{\cal O}(\partial_R^2). \label{c}
\ee
Equation (\ref{c}) and (\ref{grad}) give the response and polarization 
functions of 
finite systems in first order gradient approximation.

Now we are ready to derive approximate formulae for spherical
nuclei. In this case we can assume ${\bf q}||{\bf R}$ and we have
\be
\Pi(R,{\bf q},\omega)=\Pi^0(R,{\bf q},\omega)-{i\over q} 
\partial_R \left[1+{\omega\over
2}\partial_{\omega}\right] \Pi^0(R,{\bf q},\omega)
\label{22}
\ee
where $\Pi_0$ is the usual Lindhard polarization with spatial
dependent distributions (chemical potentials, density).
We use now further approximations. In the case
of giant resonances we are in the regime of small $q$
and ${\rm Im}\Pi^0 \sim
\omega$ such that 
\be
\left[1+{\omega\over
2}\partial_{\omega}\right] \Pi^0=i \frac 3 2 {\rm Im} \Pi^0
\label{10}
\ee
since ${\rm Re}\Pi^0\sim 1-c^2q^2/\omega^2$. 
Within the local density approximation we know that the spatial dependence is
due to the density $\rho(R)=\rho_0\Theta(R_0-R)$. Since we have for zero 
temperature 
${\rm Im} \Pi^0\propto p_f(\rho)$ we evaluate
\be
&&\partial_R {\rm Im}\Pi^0=-n_0 \delta(R_0-R)\partial_n {\rm Im}\Pi^0=
-\frac 1 3 
\delta(R_0-R) {\rm Im}\Pi^0\nonumber\\&&
\label{11}
\ee
where we assumed the density dependence carried only by the
Fermi momentum.
Now it is straightforward to average (\ref{22})
over space with the help of (\ref{11})
\be
\Pi({\bf q},\omega)&=&{3\over R_0^3} \int\limits_0^{R_0} d R R^2
\Pi(R,{\bf q},\omega)\nonumber\\
&\approx&\Pi^0({\bf q},\omega)+ i {3\over q R_0} \left[1+{\omega\over
2}\partial_{\omega}\right] \frac 1 3 \Pi^0({\bf q},\omega)
\nonumber\\
&=&\Pi^0+\Pi^{\rm surf}.
\label{3}
\ee
Consequently the surface contribution to the
polarization function reads finally with (\ref{10})
\be
\Pi^{\rm surf}({\bf q},\omega)=-{3\over 2 q R_0}{\rm Im}\Pi^0({\bf q},\omega)
\label{4}
\ee
which is real. With (\ref{3}), (\ref{4}) and (\ref{c}) we obtain
now for the structure function 
\be
S(q,\omega)=\frac 1 \pi {{\rm Im}\Pi^0\over(1-V ({\rm Re} \Pi^0+\Pi^{\rm
surf}))^2+(V{\rm Im}\Pi^0)^2}.
\label{s}
\ee
For small $q$ expansion we see that the pole of the structure function
becomes renormalized similar to what is known from the Mie mode or
surface plasmon mode \cite{osciquasi,KVO95}
\be
\omega^2={\omega_0^2\over 1-V \Pi^{\rm
surf}}.
\ee

After establishing the structure function including surface
contribution we specify the model for actual calculations. We
choose as mean field parameterization a Skyrme force (\ref{vaut}) following 
Vautherin and Brink \cite{VBR72} which leads to the isoscalar potential
\be
V=U_n-U_p=\frac{3t_0}{4}+\frac{3t_3}{8}n_0
\ee
with $t_0=-983.4$ MeV fm$^3$, $t_3=13106$ MeV fm$^6$, $x_0=0.48$ at
nuclear saturation density $n_0=0.16$ fm$^{-3}$ and
the incompressibility of $K=318$ MeV. 
Further we employ again the Steinwedel-Jensen model \cite{SJE50} 
where the basic mode inside a sphere of radius $R_0$ is given by a wave vector
\be
q_{sp}={\pi\over 2 R_0}.
\ee
This would correspond to the first order isovector mode \cite{BRV94}. 
Since this mode is spurious we have to consider the next higher
harmonics \cite{MFW99} which is
\be
q_{isgdr}={\pi\over R_0}.
\ee
The polarization function with this second order mode contains
still contributions from the spurious mode such that we have to subtract
this part \cite{D73,HSZ98}
\be
\Pi^0_{\rm ISGDR}(\omega)=\Pi^0(q_{isgdr},\omega)-\Pi^0(q_{sp},\omega).
\label{5}
\ee

\begin{figure}[h]
\parbox[]{18cm}{
\parbox[]{12cm}{
\centerline{\psfig{file=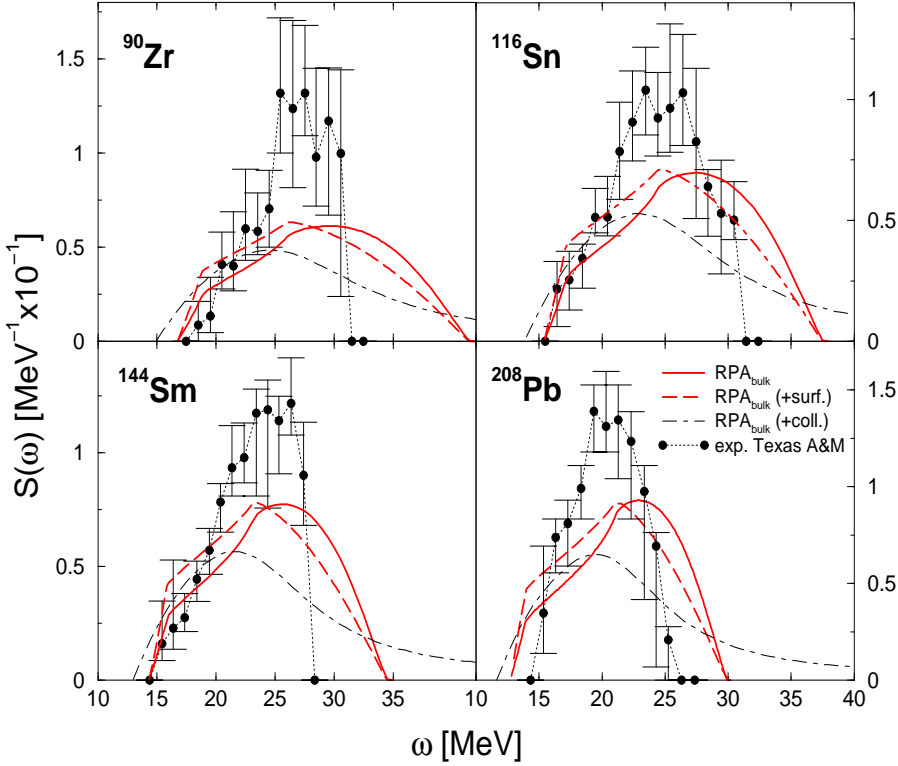,height=12cm,width=10cm,angle=-90}}}
\hspace{0.5cm}
\parbox[]{5cm}{
     \caption{The experimental structure function (T=0) 
versus theoretical values. 
The bulk RPA result (solid lines) is compared with the 
extended Thomas Fermi approximation (surface corrections, dashed lines)     
and the inclusion of collisions (dot-dashed lines). 
The latter one should be of less 
importance due to symmetry of isoscalar mode. The data suggest this case and 
support surface contributions. Circles: Normalized data from 
Ref. \protect\cite{GAR99} }}}\label{f11}
\end{figure}

In figure \ref{f11} we have plotted the 
experimental structure function together 
with different theoretical estimates according to (\ref{5}) and (\ref{s}). The 
inclusion 
of surface corrections (dashed lines) shifts the structure function 
towards the experimental 
values. The inclusion of collisions (dot-dashed lines), which should be 
wrong for 
isoscalar dipole mode due to cancellation of backscattering leads 
to really worse results.
This supports indirectly that the mode is of isoscalar
dipole type and surface dominated.

\section{Simplified model for nuclear matter situation}

The resulting wave vectors have very low values compared with the Fermi wave
vector in the Steinwedel Jensen model. This allows us to expand the Mermin 
polarization function
(\ref{mm}) with respect to small $q v_c/\omega$ ratios and $v_c$ the sound
velocity. We obtain
\beq
\Pi_a^{\rm M}(\omega)&=&\frac{\rho_a(\mu_a)}{m_a}{q^2 \over 
                       \omega (\omega+{i/\tau_a})},\label{entw}\\
\rho_a(\mu_a)&=& 2\,\lambda_a^{-3} f_{3/2}(z_a)\zeta_{corr} 
\eeq
where the thermal wave length is $\lambda_a^2=2 \pi \hbar^3/(m_a T)$,
$f_{3/2}$ the standard Fermi integral and $z_a={\rm e}^{\mu_a/T}$
the fugacity. 
The correction constant $\zeta_{corr}=1.22$ is 
introduced to fit the numerical solution of the dispersion relation (\ref{dis}) 
with the approximative expansion (\ref{entw}).
With the help of this expansion the dispersion relation
(\ref{dism}) takes the form
\beq
0&=&\bigg[\omega\Big(\omega+\frac{i}{\tau_n}\Big)-c_{nn}^2 q^2\bigg]
    \bigg[\omega\Big(\omega+\frac{i}{\tau_p}\Big)-c_{pp}^2 q^2\bigg]\nonumber\\
&&\!\!\!-\left[c_{np}^2+i{\tilde c_{np}^2 \over ( \omega +{i/\tau_n})\tau_p}
\right]\left[c_{pn}^2+i{\tilde c_{pn}^2 \over ( \omega +{i/\tau_p}) 
\tau_n}\right] q^4\label{order}
\eeq
with the partial sound velocities $c$ and $\tilde c$
\beq
c_{ab}^2=\alpha_{ab}\frac{\rho_a(\mu_a)}{m_a}, \qquad 
\tilde c_{ab}^2=\frac{T}{m_a}\frac{\frac{f_{3/2}(z_a)}{f_{1/2}(z_a)}
                                  -\frac{f_{3/2}(z_b)}{f_{1/2}(z_b)}}
                 {\frac{\tau_{ab}}{\tau_{bb}}-\frac{\tau_{aa}}{\tau_{ba}}}.  
\label{cc}
\eeq

The dispersion relation (\ref{dism}) with the dynamical relaxation times 
(\ref{mem}) is  
a polynomial of tenth (sixth) order 
corresponding to the inclusion of memory (in)dependent relaxation times. 
While most of these solutions
will just lead to parasite solutions (${\rm Re}\,\omega\le0$) we
will get two coupled modes, i.e. the isoscalar and isovector
mode. Furthermore a third mode appears at extreme asymmetries and/or strong 
collisional coupling which we will describe in the next section.

\section{New collective mode}

\begin{figure}[h]
\begin{minipage}{8.5cm}
\centerline{\psfig{file=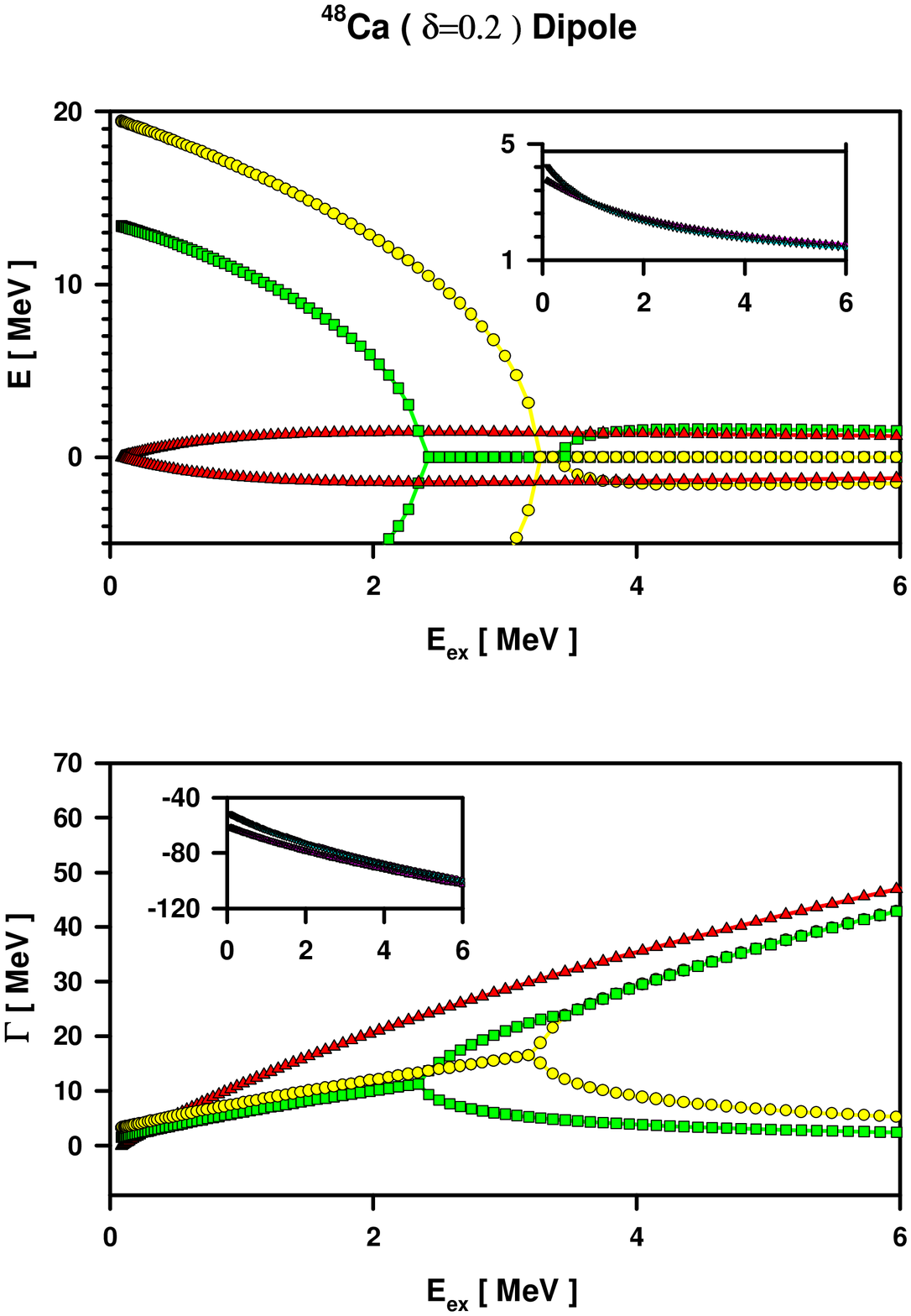,width=8.5cm,angle=0}}
\end{minipage}
\hspace{0.5cm}
\begin{minipage}{8.5cm}
\centerline{\psfig{file=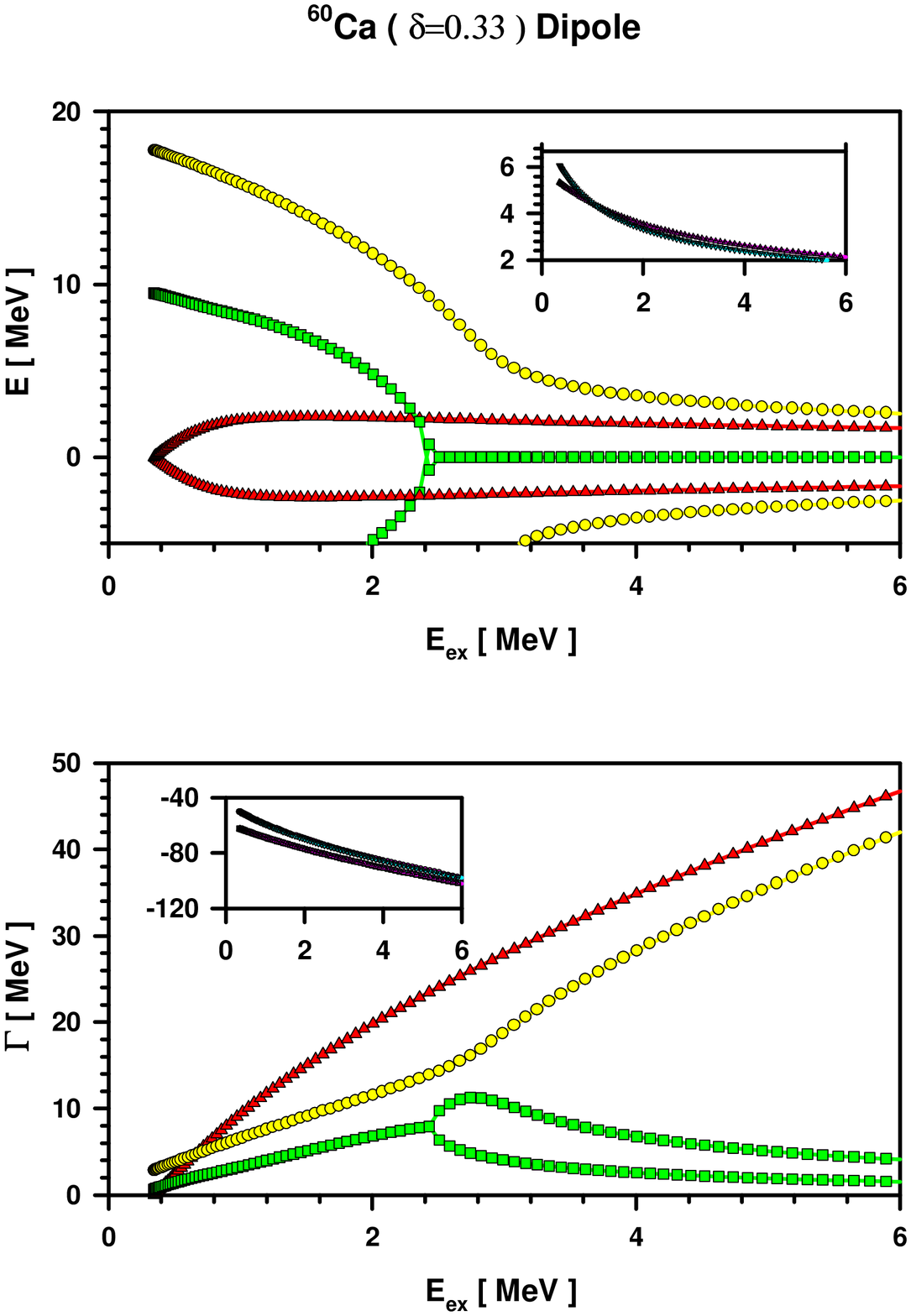,width=8.5cm,angle=0}}
\end{minipage}
 \caption{The energy and damping of the IVGDR and ISGDR of $^{48}Ca$ (left) 
          and $^{60}Ca$ (right) vs excitation energy. 
          Besides the isovector modes (circles) and 
          the isoscalar  modes (squares) 
          a soft third mode appears (triangles).\label{d2820}}
\end{figure}

Now we employ the potential (\ref{vaut}) and
assume different neutron and proton densities.
In figure \ref{d2820} we plot the isoscalar and isovector modes
versus temperature for $^{48}Ca$ with a small asymmetry $\delta=0.2$ as well
as for $^{60}Ca$ with an asymmetry $\delta=0.33$. 
The kinetic energy is
linked to a temperature within the Fermi liquid model via
Sommerfeld expansion 
\beq
{{\cal E}\over A}_{\rm kin} &=&\frac 3 5 \epsilon_f \left[
{(1+\delta)^{5/3}+(1-\delta)^{5/3}\over 2} \right .+\left . {5\over 12}\pi^2 
\left({T\over T_f}\right)^2 {(1+\delta)^{1/3}+(1-\delta)^{1/3}\over 2} \right].
\label{fer}
\eeq
This connection between temperature and excitation energy is only valid for 
a continuous Fermi liquid model. For small nuclei, the concept of 
temperature is questionable. Some improvement can be obtained by the 
definition of temperature via the logarithmic derivative of the density 
of states \cite{BM69}
\beq
T^{-1}=\frac{1}{\rho}{\partial \rho\over \partial E_{\rm ex}}
=-\frac  5 4 E_{\rm ex}^{-1}+\pi ({A\over 4 \epsilon_f E_{\rm ex}})^{1/2}
\eeq
which provides $E_{\rm ex}\approx \frac 1 4 (E/A)$ in comparison 
with (\ref{fer}) for small temperatures and small nuclei. We use this temperature to 
demonstrate possible collective bulk features in an exploratory sense. 
Of course, the surface energy and shell effects cannot be neglected for 
realistic calculations of small nuclei.

With increasing temperature the isovector and isoscalar energies decrease 
and vanish at a certain temperature. At these temperatures the damping 
becomes twofold because the damping of the spurious mode with negative 
energy becomes different from the physical mode. We can consider this 
behavior of damping as a phase transition of isospin demixture. 
At the same time a very soft mode appears due to collisional coupling which 
is only present in asymmetric matter \cite{MWF97}. This mode is more 
pronounced in the next figure \ref{d2820} for $^{60}Ca$ with an asymmetry 
of $\delta=0.33$.

We see that the isovector mode does not disappear 
but turns over into a flat decrease 
with increasing temperature. This behaviour is coupled with the pronounced 
soft mode. In comparison with the more symmetric $^{48}Ca$ we see a different 
behavior of the damping where the isoscalar mode vanishes. 
Also the isovector mode appears 
unique and not two-fold. Now a clear transition of damping behavior for the 
isovector mode is recognizable 
which can be considered as a transition from zero to first sound damping.

Besides the standard isovector and isoscalar modes
we observe a
build up of a very soft mode with a centroid energy around $1$ MeV.
This mode appears due to the collisional coupling $\bar c_{ab}^2$ of (\ref{cc}).
When we turn off the relaxation times, i.e. the collision integral,
this mode is vanishing as well as in symmetric nuclear matter, see discussion after 
(\ref{D}). It shows that this mode appears due to
collisional coupling of isovector and isoscalar modes. The
corresponding damping of the crossed mode is continuously increasing with temperature.

One may argue whether this third mode
can really appear in the system. A simple consideration may
convince us about the possible existence of such mode. Let us
assume a coupled set of two type of harmonic oscillators (neutrons
and protons) interacting between the same sort of particles with strength $k_n$
and $k_p$, respectively and between different sorts with $k_{np}$.
Let us choose for simplicity only two neutrons and two protons.
Then
we obtain the coupled system of harmonic oscillators with
frequencies $\omega_n^2=k_n/m$, $\omega_p^2=k_p/m$ and
$\omega_{np}^2=k_{np}/m$. The solution yields
three basic modes in the system, i.e.
$\omega^2=2(\omega_n^2+\omega_{np}^2), 4 \omega_{np}^2,
2(\omega_p^2+\omega_{np}^2)$. If we neglect the different coupling
between neutrons and protons $\omega_{np}$ we only obtain
two modes analogously to isovector and isoscalar ones. We see
that the coupling between neutrons and protons can lead to the
appearance of a third mode.

Let us now compare the found new mode with the experimental evidence.
There are some hints for a soft mode in $^{11}Be$ \cite{C96}.
The authors have observed a low lying structure at around
$6$ MeV excitation energy with a damping of around $1$ MeV which has not been
reproduced yet even within refined coupled channel calculations
\cite{EBS95}. A standard explanation would give as the origin a weakly-bound single 
particle neutron orbital. 
The observed broad structure at $6$ MeV might
be explained alternatively as the presented new coupled mode.
The centroid energy as well as damping width at least seem to
suggest this interpretation.

\chapter{Nonlinear effects beyond linear response}\label{oct}

While most of the theoretical treatments of oscillations rely on the linear response 
method or RPA methods, large amplitude oscillations require methods beyond.  
In particular the question of the appearance of chaos has
recently been investigated \cite{BGZS94,schuckbaldo,Mr97}. 
The hypothesis was established that the octupole mode is over-damped due to negative 
curved surface and consequently additional chaotic damping
\cite{JarSwi93,BloShi93,BloSka97}. Here we want to discuss in which conditions one 
might observe octupole modes at least in Vlasov - simulation of giant resonances.
We will consider different initial conditions of isoscalar giant 
resonances and will demonstrate that the appearance of octupole modes is dependent on 
the initial configuration which in turn demonstrates the nonlinear behavior beyond 
linear response.

\begin{figure}[h]
\parbox[]{18cm}{
\parbox[]{10cm}{
\psfig{file=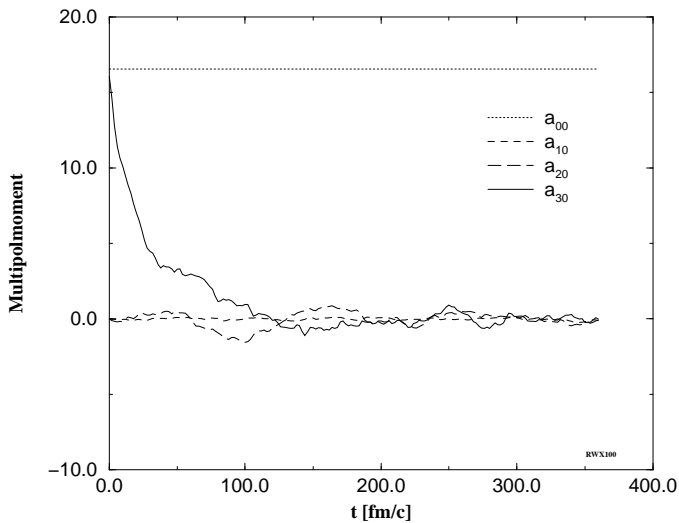,width=9cm}}
\hspace{0.5cm}\parbox[t]{7cm}{
\caption{The picture shows the time evolution of multipole moments
  $\protect\sqrt{\frac{2l+1}{4\pi}} a_{l0}(t), l=0,1,2,3$ of mass
  density concerning to normalization, dipole, quadrupole and octupole
  oscillation. Excitation is due to
 modification in momentum space only.\label{X100}} 
}}
\end{figure}

\begin{figure}[h]
\parbox[]{18cm}{
\parbox[]{10cm}{
\psfig{file=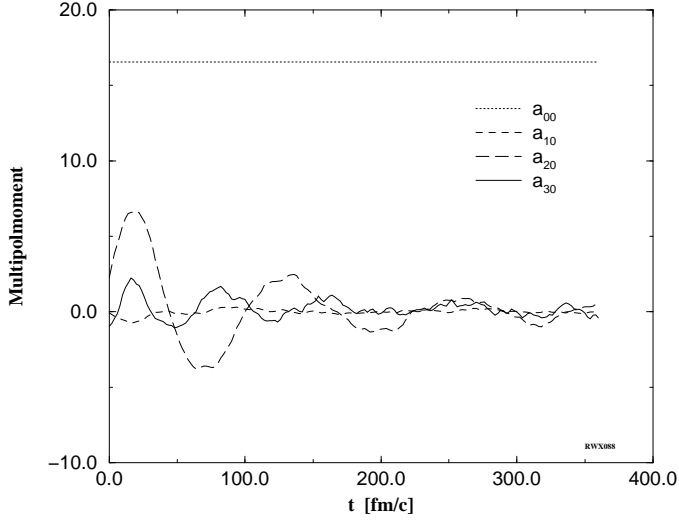,width=9cm}}
\hspace{0.5cm}\parbox[t]{7cm}{
\caption{This picture shows the time evolution of multipole moments
  $\protect\sqrt{\frac{2l+1}{4\pi}} a_{l0}(t), l=0,1,2,3$ of mass
  density. Excitation is due to
 asymmetric splitting (0.3 to 0.7) of the nucleus in two pieces using
 a plane in coordinate space. Then both pieces are accelerated towards each
 other.\label{X088}}}}
\end{figure}

As a first initial condition we use the ground state distribution of coordinates while 
the momentum distribution is deformed anisotropically.
We have modified the momenta in a way which corresponds to a giant octupole mode. The 
local densities and currents remain the same as in ground state.
Figure \ref{X100} shows that at start time $t=0$ there is a pure
giant octupole, which is
damped out and a quadrupole resonance develops instead. The monopole and dipole 
amplitudes which should remain constant document the stability of simulation.
In agreement with the already mentioned hypothesis, the
octupole mode is over-damped. The figure shows the nonlinear behavior of mode 
coupling. Within the linear response the damping rate is expected to be independent of 
initial conditions. We choose now other initial conditions to show that the result is 
very much dependent on initial conditions. 
Therefore we
split the nucleus in two parts of mass ratio 3:7 in accordance with
the symmetry of the 
octupole oscillation 
and accelerate both pieces towards each other. Experimentally this
might be realized as a central collision of two nuclei with
corresponding masses.
In figure \ref{X088} a clear quadrupole resonance appears and also 
a smaller octupole resonance can be seen. Both are damped out.
Consequently there is no evidence for an over-damped octupole mode in this case.

In order to
understand the different initializations, we split
the kinetic energy into a thermal part and a collective part
according to 
$\langle {\bf p}^2\rangle=\left\langle({\bf p}-\langle {\bf p}\rangle)^2\right\rangle 
+
\langle {\bf p}\rangle^2$.
We analyze the time
development of the total collective energy in the system
\begin{eqnarray}
E_{coll}(t) &=& \frac{h^2}{2m} \int d{\bf r} \varrho({\bf r}) \langle {\bf 
p}\rangle^2({\bf r},t)
\label{ecol}
\end{eqnarray}
with the mean current
\begin{eqnarray}
\langle {\bf p}\rangle({\bf r},t) &=& \frac{1}{\varrho({\bf r})} \int \frac{d{\bf 
p}}{(2\pi)^3} \;{\bf p}
\; f({\bf p,r},t)
\end{eqnarray}
and the density
\begin{eqnarray}
\varrho({\bf r},t) &=& \int \frac{d{\bf p}}{(2\pi)^3} \; f({\bf p,r},t) \; .
\end{eqnarray}

\begin{figure}[h]
\parbox[]{18cm}{
\parbox[]{10cm}{
\psfig{file=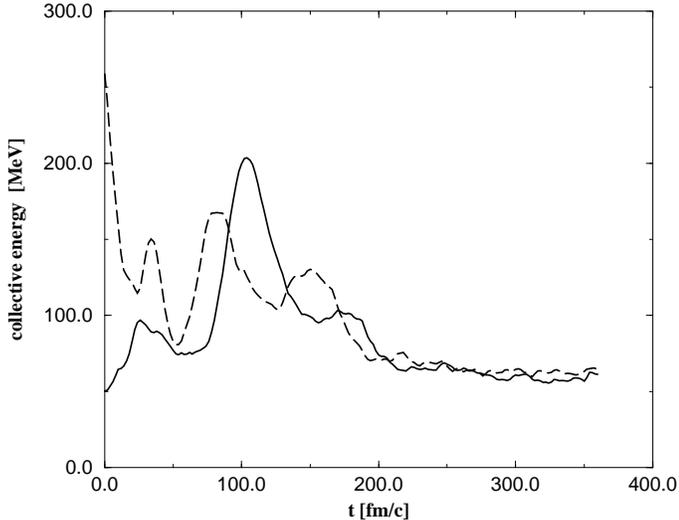,width=9cm}
}
\hspace{0.5cm}\parbox[t]{7cm}{
\caption{The time evolution of collective energy. The solid line
 corresponds to the excitation scheme of figure \ref{X100} and the
 long dashed line corresponds to the figure
  \ref{X088}, respectively. While the first starts without collective
  energy, the second one starts with maximal collective energy.
\label{coll}}
}}
\end{figure}

In figure \ref{coll} the development of collective energy can be
seen. There is a background of about 50~MeV due to fixed correlations
caused by finite width of pseudo-particles 
 as one can see from the following estimation. Using (\ref{2}) and (\ref{3}) in 
(\ref{ecol}) one obtains
\begin{eqnarray}
E_{coll}
   &\approx& {\textstyle \frac{1}{\varrho_0}} \sum_{i=1}^{AN} \sum_{j=1}^{AN}
   {\textstyle \frac{1}{N^2}} p_i p_j \int dr f_S(r-r_i(t),\sigma_r)
   f_s(r-r_j(t),\sigma_r)
\nonumber\\
&\approx& {\textstyle \frac{1}{\varrho_0}} \sum_{i=1}^{AN} \sum_{j=1}^{AN}
   {\textstyle \frac{1}{N^2}} p_i p_j f_S(r_i(t)-r_j(t),\sqrt{2}\sigma_r)  \nonumber\\
&\ge& {\textstyle \frac{1}{\varrho_0 N}
\frac{1}{(\sqrt{2\pi}\sqrt{2}\sigma_r)^3}} \sum_{i=1}^{AN} {\textstyle
\frac{1}{N}}
p_i.
\end{eqnarray}
For simulation parameters of $^{208}Pb$, 
$\rho_0=0.162\; {\rm fm}^{-3}$, $N=75$, 
$\sigma_r=0.53\;{\rm fm}$,
we obtain a basic collective energy of ${54 \rm MeV}$. Using more test-particles would 
diminish this level.

The solid line corresponding to
figure \ref{X100} shows no initial
collective energy. 
The exclusive initial excitation in momentum space without correlation
in the spatial domain leads to zero initial collective energy.
These correlations are forming during time evolution.
Of course, there is no  center of mass motion, otherwise we would see
just the mean streaming velocity.
 
This situation is changed if we use the second preparation with simple
momentum--space--correlations. 
The long dashed line in figure
\ref{coll} shows initial collective correlations corresponding to
figure \ref{X088}.
Since we can deposit enough collective energy in this case, we
observe a clear octupole motion.

In order to compare with \cite{BloShi93} we calculate the adiabaticity
index $\eta$
defined in \cite{BloShi93} as the ratio of maximum radial
surface velocity to the maximum particle speed. A smaller ratio denotes a
more adiabatic shape changes in relation to the particle speed.
In analogy we define such index as a ratio
\begin{eqnarray}
\eta &=& \frac{\frac{\partial}{\partial t}\sqrt{<r^2>(\vartheta)}}{v_F}
\end{eqnarray}
of the root mean square radius speed in forward direction (opening angle
$\vartheta$) and the Fermi velocity.
With opening angle 0.4 rad we obtain a maximum $\eta=0.12$ for figure \ref{X100} and
$\eta=0.30$ for figure \ref{X088}.
This shows that we are essentially still in the adiabatic regime described in
\cite{BloShi93}. 

The nonlinear behavior described so far already documents that we are in a regime of 
large amplitude oscillations where linear response fails. The corresponding radius 
elongation
in coordinate space varies about 10 \%.

\chapter{Summary}

We have investigated the giant resonance oscillations by solving the
collision-less Vlasov equation as well as by linear response theory.

With the help of numerical solution of the Vlasov equation we could
reproduce the mass dependence of energy with one fixed parameter of test-particle
width for both giant monopole and giant dipole modes. A multipole
analysis was performed which has allowed us to characterize the
corresponding excitation.
It was shown, that the asymmetry influences the collective behavior.
With increasing asymmetry the energy decreases while the damping increases.
The damping due to finite systems amount to $2$MeV
almost constant for all mass numbers which
underestimates the experimental values considerably. This motivates to
search for additional damping mechanism which was found to be due to the collisional 
correlations.

The linear response
can be used in a simplified liquid drop model to describe the giant
resonances in asymmetric matter. We find that the collisional
contribution as well as surface contributions are both important to
reproduce the experimental damping of giant dipole resonances. While
for
ground state resonances it is sufficient to add the damping of finite
size effects from simulation with the collisional contribution from
linear response theory, for the temperature behavior we have needed
also the deformation of the surface. The combined model between
surface and collisional contribution, weighted properly due to their
collision frequencies, is able to reproduce the experimental damping
curve with temperature as well as the structure function.
A higher order mode like the recently measured ISGDR mode has been described within this simple linear response model.

We have observed that due to correlational coupling
there can exist a new mode which appears besides isovector and
isoscalar modes in asymmetric nuclear matter due to collisions. 
We suggest that
this mode may be possible to observe as a soft collective
excitation in asymmetric systems. The transition from zero sound damping to 
first sound damping behavior should become possible to observe for isovector 
modes since they do not vanish at this transition temperature like in 
symmetric matter. At a certain critical temperature the collective isoscalar mode 
vanishes and the damping becomes two-fold. This can be considered as a phase 
transition of demixture.

On the example of octupole resonances 
in finite nuclei the dependency on
the initial configuration is demonstrated.
The appearance of an octupole mode was shown by 
correlating the spatial and momentum
initial excitation.
It is possible to
excite an octupole mode with sufficient collective energy deposited initially.
We suggest that isoscalar giant octupole resonances should be
possible to observe in nuclear collisions of mass ratio about 3:7 corresponding to 
octupole symmetry. This effect illustrates a mode beyond linear response.

\section{Acknowledgment}

The authors are obliged to M. Vogt who has contributed the calculation
of the Lyapunov exponent. M. DiToro is thanked for numerous
discussions and the LNS Catania where part of this work was done 
for hospitality. Valuable hints and comments by J. D. Frankland
(GANIL) are also gratefully acknowledged.

\appendix

\chapter{Derivation of non-Markovian relaxation time}\label{ap}

Here we shortly sketch the derivation of the damping of
a collective mode within a Fermi gas and a Fermi liquid model. 
We will show that we get the latter one from the Fermi gas model
with an additional contribution from the quasi-particles.

We start with the Fermi gas where the dispersion relation between
momentum and energy is given by $\epsilon=p^2/2m$ and will show
later what has to be changed for a Fermi liquid where $\epsilon$
is a solution of the quasiparticle dispersion relation. We will
see that the contributions from the quasi-particles alone leads to the Landau
formula of zero sound damping \cite{AKH59,LL79}
\be
\gamma\propto \left[1+\left({\Omega\over  2\pi T} \right)^2\right]
\label{lan}.
\ee

Our considerations use conveniently the Levinson equation
for the reduced density matrix $f$ which is valid at short time processes
compared to inverse Fermi energy $\hbar/\epsilon_f$ and which collisional side has the 
form:
\be
I_1(t)={2 g\over \hbar^2}\int\limits_0^{\infty} 
d \tau \int {d{\bf p}_2 d{\bf p}_3
d{\bf p}_4\over (2 \pi \hbar)^6}|T|^2
\cos{\left(\int\limits_t^{t-\tau}\Delta\epsilon(\tau)
d \tau/\hbar\right)}\delta(\Delta {\bf p})
\left({\bar f_1}{\bar f_2}f_3f_4-f_1f_2{\bar f_3}{\bar f_4}\right)_{t-\tau}
\label{levin}
\ee
where ${\bar f}=1-f$, $\Delta p=p_1+p_2-p_3-p_4$ etc., g is the
spin-isospin degeneracy and the transition probability is given by
the scattering $T$-matrix. In case that the quasiparticle 
energies $\epsilon(t)$ become time independent like in the Fermi gas model, 
the integral in the $cos$ function reduces to the familiar 
expression $\Delta \epsilon \tau$. We linearize this collision integral
with respect to an external disturbance according to
\be
f=n+\delta f \label{lineari}
\ee
where n is the equilibrium distribution. 
Clearly two contributions have to be distinguished, 
the one from the quasiparticle energy and the one from 
occupation factors \cite{FMW98}. 
First we concentrate on the Fermi gas model where we have only 
the contribution of the occupation factors and will later add the 
contribution of the quasiparticle energies for Fermi liquid model. 
We obtain after Fourier transform of the time
\be
\delta I_1&=&\Big\langle \frac{\hbar}{2} \left[\delta_+(\Delta
\epsilon+\Omega)+\delta_-(\Delta\epsilon-\Omega)\right]
\left(\delta
F_1+\delta F_2-\delta F_3-\delta F_4\right)(\Omega)\Big\rangle.
\label{abb1}
\ee
Here we use the abbreviation
\be
\langle ...\rangle&=&{2 g\over \hbar^2} \int {d{\bf p}_2 d{\bf p}_3
d{\bf p}_4\over (2 \pi \hbar)^6}|T|^2 \delta(\Delta p)...={m^3 g\over \hbar^2 (2 \pi 
\hbar)^6}\int d\epsilon_2
d\epsilon_3d \epsilon_4 \int {d \phi \sin \theta d\theta d \phi_2
\over \cos (\theta/2)} |T|^2...\nonumber\\&&
\label{int}
\ee
where the last line appears from standard integration techniques
at low temperatures.
Further abbreviations are
\be
\delta_{\pm}(x)&=&\pi \delta(x)\pm i{{\cal P}\over x}\approx \pi
\delta(x)\nonumber \\
\delta F_1&=&-\delta f_1({\bar n_2}n_3 n_4 +n_2 {\bar n_3}{\bar
n_4}).
\ee
The approximation used in the first line consists in the neglect
of the off-shell contribution from memory effects. This is
consistent with the used integration technique (\ref{int}). This
terms would lead to divergences which has to be cut off \cite{MK97}.

Neglecting the backscattering terms $\delta F_{2/3/4}$ we obtain
from (\ref{abb1}) a relaxation time approximations with the
relaxation time
\be
{1\over \tau(\epsilon_1)}={3\over 4 \pi^2 \tau_0}
\int\limits_{-\lambda}^{\infty} d x_2 dx_3 dx_4 \left[\delta (\Delta
x+\omega)+\delta(\Delta x-\omega)\right]
({\bar n_2} n_3 n_4+n_2 {\bar n_3}{\bar n_4})
\label{int1}
\ee
with $\omega =\Omega/T$, $x=(\epsilon-\mu)/T$, $\lambda =\mu/T$
and the time
\be
{1\over \tau_0}={2 g m T^2 \over 3 \hbar^3} \sigma.
\ee
Here we have used the definition of cross
section $|T|^2=(4 \pi\hbar^2/m)^2d\sigma/d\Omega$ and have
assumed a constant cross section $\sigma$.
To calculate (\ref{int1}) one needs the standard integrals for
large ratios of chemical potentials $\mu$ to temperature $\lambda=\mu/T$
\be
&&\int\limits_{-\lambda}^{\infty} d x_2 dx_3 dx_4 n_2 {\bar
n_3}{\bar n_4} \delta(\Delta x \pm \omega)=\frac 1 2 {\bar
n_1}(x_1\pm\omega)\left[\pi^2+(x_1\pm\omega)^2\right]
\label{stand}
\ee
to obtain
\be
{1\over \tau(\epsilon_1)}= {3 \over 8 \pi^2 \tau_0} \left[2
\pi^2+(x_1+\omega)^2+(x_1-\omega)^2\right].
\label{t1}
\ee
Further we employ a thermal averaging in order to obtain the mean
relaxation time finally\footnote{One has to use the identities
valid up to $o(\exp[-\lambda])$
\[
\int\limits_{-\lambda}^{\infty} d x n {\bar n}=1; \qquad
\int\limits_{-\lambda}^{\infty} d xx^2 n {\bar n}={\pi^2\over 3}.
\]}
\be
{1\over \tau_{\rm gas}}=\int\limits_{-\lambda}^{\infty} d x_1 n_1 {\bar
n_1} {1\over \tau(\epsilon_1)}={1\over \tau_0}\left[1+3\left({\omega\over 2
\pi}\right)^2\right].
\label{avt}
\ee
If we do not use the thermal averaging but take 
(\ref{t1}) at the Fermi energy $\epsilon_1=\epsilon_f$ we will obtain
\be
{1\over \tau(\epsilon_f)}={1\over \tau_0}\left[\frac 3 4 +3\left(
{\omega\over 2\pi}\right)^2\right].
\label{tep}
\ee
We see that both results disagree with the Landau result of
quasiparticle damping (\ref{lan}) by factors of 3 at different
places \cite{FMW98,AB92,KMP93,MKH95}. 
We have point out that the result at fixed Fermi energy will
lead to unphysical results for the Fermi liquid case. Therefore
we consider the thermal averaged result as the physical one.

We now turn to the Fermi liquid model and replace the free
dispersion $\epsilon=p^2/2m$ by the quasiparticle energy
$\epsilon_p$. Than the variation of the collision integral gives
an additional term which comes from the time dependence of the
quasiparticle energy on the $\cos$-term of (\ref{levin}). We have
instead of the sum of two complex conjugate exponentials in
(\ref{levin}) an additional contribution from the linearization
of the exponential
\be
\delta \exp{\left (i\Delta\int\limits_t^{t-\tau} d {\bar t }
\epsilon({\bar t}) \right )}&=&{\rm e}^{-i\Delta \epsilon \tau}
\left
(1-i \Delta
\int\limits_t^{t-\tau} d{\bar t}[\epsilon({\bar
t})-\epsilon]
\right )
-{\rm e}^{-i\Delta \epsilon \tau}
\nonumber\\
&=&{i T\over \hbar} {\rm e}^{-i\Delta \epsilon \tau/\hbar}
\Delta\int\limits_t^{t-\tau}
d{\bar t } {\delta f({\bar t})\over n {\bar n}}.
\ee
In the last line we have replaced the variation in the
quasiparticle energy $\epsilon(t)-\epsilon$
by the variation in the distribution
function $\delta f$ due to the identity \cite{LL79}
\be
\delta
f(t)&=&f(t)-n(\epsilon)=f(t)-n(\epsilon(t))-[n(\epsilon)-n(\epsilon(t))]
\nonumber\\
&\approx &-n'(\epsilon(t)-\epsilon)={n{\bar n}\over
T}\left[\epsilon(t)-\epsilon\right]
\ee
where we assumed within the quasiparticle picture that
$f(t)=n(\epsilon(t))$. This leads now to an additional part in
the relaxation time which we write analogously to (\ref{abb1})
\be
{1\over \tau_c(\epsilon_1)}&=&\left\langle \frac{\hbar}{2} {\delta_+(\Delta
\epsilon-\Omega)-\delta_+(\Delta\epsilon+\Omega)\over
\Omega} {{\bar n_1}{\bar n_2}n_3 n_4-n_1 n_2 {\bar n_3}{\bar
n_4}\over n_1 {\bar n_1}} \right\rangle.
\label{abb2}
\ee
Using again (\ref{stand}) we obtain
\be
{1\over \tau_c(\epsilon_1)}={-3\over 4 \pi^2\tau_0}
\Bigg \{ \frac{{\bar n}(x_1+\omega)\left[\pi^2+(x_1+\omega)^2\right]}
{2\omega({\rm e}^{-\omega}-1)}
+\,[\omega \leftrightarrow - \omega] \Bigg\}  
\label{tauepsq}
\ee
and get after thermal averaging (\ref{avt})\footnote{Here one uses
[$o(\exp[-\lambda])$]
\[
\int\limits_{-\lambda}^{\infty} dx n=\lambda;\quad
\int\limits_{-\lambda}^{\infty} dx x n=-\frac 1 2 \lambda^2+\frac
1 6 \pi^2;\quad
\int\limits_{-\lambda}^{\infty} dx x^2 n=\frac 1 3 \lambda^3
\].}
\be
{1\over\tau_c}={1\over \tau_0}\left[1+
\left({\omega\over 2 \pi}\right)^2\right].
\label{avt1}
\ee
Taking instead of thermal averaging the value at Fermi
energy $(\epsilon_1=\epsilon_f)$ in (\ref{tauepsq}) we find
\be
{1\over \tau_c(\epsilon_f)}={3 (\pi^2+\omega^2)\over 2
\pi^2\tau_0\omega} {{\rm e}^{\omega}-1\over {\rm e}^{\omega}+1}.
\label{tep1}
\ee
Here we like to point out that the Landau result (\ref{lan}) 
appears in (\ref{avt1}) (see also in Ref. \cite{AB92,ABB95,AYG98,KMP93,MKH95}).

Adding now (\ref{avt}) and (\ref{avt1}) we obtain a final relaxation time for
the Fermi liquid model
\be
{1\over \tau_{\rm liq}}={2\over \tau_0} \left[1+2 \left({\omega\over 2
\pi}\right)^2\right].\label{avtliq}
\ee
which is the main result in this paper.
It contains the typical Landau result of zero sound (\ref{lan}) except 
the factor 2 in front of the frequencies.
Comparing (\ref{avtliq}) with the Fermi gas model (\ref{avt}) we
see that in the limit of vanishing temperature the Fermi liquid value is
lower with $\propto 2 \Omega^2$ compared to the Fermi gas $\propto 3
\Omega^2$. 
Further for vanishing frequencies (neglect of memory effects)
the Fermi liquid model leads to twice the relaxation rate than the Fermi gas
model.  The coefficient of temperature increase is than twice
larger for the Fermi liquid than for the Fermi gas. 
If we consider the relaxation times at Fermi energy (no thermal averaging)
(\ref{tep}) and (\ref{tep1}) we find the same results as above 
in the limit of vanishing frequencies. For vanishing temperature 
only the Fermi gas (\ref{tep})  coincides  with the result of (\ref{avt})
$\propto 3\Omega^2$. Expression (\ref{tep1}) goes to zero for $T=0$
and underlines the necessity to thermal average the value.


\end{document}